\documentclass[12pt]{article}

\def\ahat{{\hat{\bf a}}}
\def\Avec{{\bf A}}
\def\evec{{\bf e}}
\def\Jvec{{\bf J}}
\def\Uvec{{\bf U}}
\def\vhat{{\hat{\bf v}}}
\def\Vvec{{\bf V}}
\def\nhat{{\hat{\bf n}}}
\def\zhat{{\hat{\bf z}}}
\def\min{\mathop{\rm min}\nolimits}
\def\max{\mathop{\rm max}\nolimits}
\def\sign{\mathop{\rm sign}\nolimits}

\begin{document}
\title{Uniform Semiclassical Approximation for the Wigner $6j$ Symbol in 
Terms of Rotation Matrices}

\author{Robert G. Littlejohn\\
Liang Yu\\
University of California\\
Berkeley, CA 94720-7300 USA}
\maketitle
\begin{abstract}
  A new uniform asymptotic approximation for the Wigner $6j$ symbol is
  given in terms of Wigner rotation matrices ($d$-matrices).  The
  approximation is uniform in the sense that it applies for all values
  of the quantum numbers, even those near caustics. The derivation of
  the new approximation is not given, but the geometrical ideas
  supporting it are discussed and numerical tests are presented,
  including comparisons with the exact $6j$-symbol and with the
  Ponzano-Regge approximation.
\end{abstract}

\section{Introduction}
\label{intro}
The Wigner $6j$-symbol is used in the recoupling of three angular
momenta, and finds many applications in atomic, molecular and nuclear
physics.  These are explained in standard references on angular
momentum theory\cite{Edmonds60, BiedenharnLouck81, BrinkSatchler93, 
Varshalovichetal81}.  For example, the $3nj$-symbols and their
asymptotic properties are central to certain algorithms for
the calculation of scattering amplitudes in three-body interactions
(De~Fazio et al\cite{DeFazioetal03}, Anderson and
Aquilanti\cite{AndersonAquilanti06}).  These methods make use of the
relationship between the $3nj$-symbols and discrete orthogonal
polynomials (Aquilanti et
al\cite{Aquilantietal95,Aquilantietal01a,Aquilantietal01b} and
references therein). 

The $6j$-symbol possesses a remarkable semiclassical approximation,
first obtained by Ponzano and Regge\cite{PonzanoRegge68} through some
inspired guesswork, that is linked in a highly symmetrical manner to
the geometry of a tetrahedron in three-dimensional space. This formula
was first proven rigorously by Schulten and
Gordon\cite{SchultenGordon75}, who also gave practical means of
computing the $6j$-symbol by recursion relations.  More recently the
$6j$-symbol has attracted attention for the role it plays in quantum
gravity, which has led to more geometrical treatments of its
asymptotic properties.  References in this area include
Roberts\cite{Roberts99} and Charles\cite{Charles08}.

The formula of Ponzano and Regge has the usual properties of primitive
semiclassical approximations, for example, it diverges at the
classical turning points (the caustics).  Since the $6j$-symbol is
only defined for discrete values of the quantum numbers, it is
unlikely to fall exactly on a caustic, but it can come very close, and
the Ponzano-Regge formula does not provide a good approximation for
such values.  Thus there is interest in uniform approximations that do
not suffer from caustic singularities and that are valid over as wide
a range of quantum numbers as possible.

In addition to proving the Ponzano-Regge formula for the $6j$-symbol,
Schulten and Gordon also provided uniform asymptotic approximations of
the Airy-function type that are valid in a region passing through a
turning point.  The $6j$-symbol, however, has two turning points when
one of the $j$'s is varied and the others held fixed, in a manner
reminiscent of an ordinary oscillator in one dimension.  The Airy-function
uniform approximation can only cover one of these at a time, and if
two Airy-function approximations are used, then they do not match
smoothly in the middle.  This suggests that a uniform
approximation of the Weber-function type (one based on harmonic
oscillator wave functions) should be used, that would cover both
turning points at once.

It turns out, however, that this cannot be done, in general.  An idea
of why this is so is given by Fig.~\ref{noweber}, which shows the
$6j$-symbol and the Ponzano-Regge approximation as a function of
$j_{12}$ for certain values of the other five $j$'s (notation is
explained by (\ref{normal6j}) below).  The sticks in the diagram give
the exact values of the $6j$-symbol, while the curve is the
Ponzano-Regge approximation.  Although the $6j$-symbol is only defined
for discrete values of $j_{12}$, the Ponzano-Regge approximation is
defined for continuous values of the parameters, and allows us to talk
of the phase of the $6j$-symbol in an unambiguous manner.  The
Ponzano-Regge approximation shows the expected divergences at the
caustics or turning points, and the Airy function behavior at the
right turning point.  In the classically allowed region between the
turning points the wave length decreases as we move to the left,
corresponding to an increase in the variable conjugate to $j_{12}$
(the angle $\phi_{12}$, defined in Sec.~\ref{phasespace6j}below).  At
the left turning point the Airy function is not so obvious, because it
is multiplied by a rapidly oscillating cosine term.  This is because
the angle $\phi_{12}$ conjugate to $j_{12}$ reaches the value
$\pm\pi$ at the right turning point.

The behavior illustrated in Fig.~\ref{noweber} cannot be matched by a
Weber function (harmonic oscillator eigenfunction), for which both
turning points have the behavior of an unmodulated Airy function.
Related to this is the fact that the difference between the
Ponzano-Regge phases at the two turning points is not of the form
$(n+\frac{1}{2})\pi$ for the parameters shown in Fig.~\ref{noweber}.
For harmonic oscillator eigenfunctions the difference in the action
between the two turning points is always of the form
$(n+\frac{1}{2})\pi$, a condition that is equivalent to the
single-valuedness of the WKB wave function (that is, it is the
Bohr-Sommerfeld quantization condition).  One can say that the
$6j$-symbol has a nonstandard matching condition of the two WKB
branches at the lower turning point for the parameters in
Fig.~\ref{noweber}.

Ultimately there are topological reasons for the failure of the Weber
function as a standard form for a uniform approximation in cases such
as that illustrated in Fig.~\ref{noweber}.  Uniform approximations are
based on a smooth, area preserving map between the phase space of the
given problem and the standard problem.  But the phase space of the
$6j$-symbol is a sphere (we call it the $6j$-sphere), and the phase
space of the harmonic oscillator is a plane.  These two spaces cannot
be continuously mapped into one another.

We have noticed, however, that the phase space that arises in the
semiclassical analysis of the Wigner $d$-matrices (rotation matrices)
is also a sphere.  We call this the $d$-sphere, to distinguish
it from the $6j$-sphere.  Moreover, the classical orbits on the two
spheres are topological circles in both cases, and in both cases
intersecting orbits always intersect in two points, unless they are
tangent, in which case there is one intersection point.  
There are also cases in which the orbits do not intersect at all,
corresponding to classically forbidden regions.  As is well known, the
classical orbits and their intersections provide the geometrical
framework for the construction of semiclassical (asymptotic)
approximations.  These topological features are the same for both
the $6j$-sphere and the $d$-sphere, suggesting that one can be mapped
into the other by a smooth transformation that takes a pair of orbits on
one sphere into the pair of orbits on the other.  

We have worked out the details of this mapping and the corresponding
uniform approximation.  The resulting approximation is smooth and
uniform over the entire range of quantum numbers $j_{12}$ and $j_{23}$
(for fixed $j_1$, $j_2$, $j_3$ and $j_4$).  For most values of the
$j$'s it is more accurate than the Ponzano-Regge approximation,
certainly near the caustics but also at most other places.  We have
found no places where it is dramatically worse than the Ponzano-Regge
approximation.

In this paper we shall present the new uniform approximation itself,
as well as some of the geometric rationale behind it, which helps
considerably in understanding the formula and the various regions that
it covers.  In addition, we shall present some numerical tests of the
new formula and comparisons with the exact $6j$-symbol and the
Ponzano-Regge approximation.  We shall not, however, present the
details of the derivation.

The outline of this paper is as follows.  In
Sec.~\ref{background6j} we present a collection of facts about the
$6j$-symbol and its asymptotic or semiclassical approximation (the
Ponzano-Regge formula), emphasizing the spherical phase space and its
geometrical ramifications.  In Sec.~\ref{dmatrices} we present a
geometrical treatment of the asymptotic properties of the rotation
matrices ($d$-matrices) that emphasizes the similarities with the
$6j$-symbol.  In Sec.~\ref{unifappx} we outline the ideas behind the uniform
approximation of the $6j$-symbol in terms of $d$-matrices, we present
the actual formulas, and we present some numerical tests.  Finally, in
Sec.~\ref{conclusions} we make some conclusions.

\section{The $6j$-symbol}
\label{background6j}

\subsection{Quantum Mechanics of the $6j$-symbol}
\label{qm6j}

We set $\hbar=1$, so all angular momenta are dimensionless.  We label
the $j$'s in the $6j$ symbol by
	\begin{equation}
	\left\{ \begin{array}{ccc}
	j_1 & j_2 & j_{12} \\
	j_3 & j_4 & j_{23}
	\end{array}\right\},
	\label{normal6j}
	\end{equation}
which is how it would used when recoupling three angular momenta.

The quantum number $j_i$, $i=1,2,3,4,12,23$, just gives the magnitude
of the angular momentum, and does not specify the sign of the
operator.  For example, instead of coupling three angular momenta to
obtain a fourth, that is, setting $\Jvec_4 = \Jvec_1+\Jvec_2+\Jvec_3$,
we can couple four angular momenta with a sum of zero,
	\begin{equation}
	\Jvec_1 + \Jvec_2 + \Jvec_3 + \Jvec_4 =0
	\label{Jsumvanishes}
	\end{equation}
(effectively changing the sign of $\Jvec_4$).  This is how we shall
regard the recoupling problem in this paper.  Usually we will
think of $j_i$, $i=1,2,3,4$ as given, while $j_{12}$ and $j_{23}$ are
variable intermediate angular momenta that result from the coupling of
the first four.  They are the quantum numbers of the squares of the
operators
	\begin{equation}
	\Jvec_{12} = \Jvec_1 + \Jvec_2, \qquad
	\Jvec_{23} = \Jvec_2 + \Jvec_3.
	\label{J12J23defs}
	\end{equation}

With this interpretation, the $6j$-symbol in the form (\ref{normal6j})
is proportional to the unitary matrix element $\langle j_{12} \vert
j_{23} \rangle$ that takes one from the eigenbasis of one intermediate
angular momentum ($j_{12}$) to the eigenbasis of the other ($j_{23}$).
These bases span the subspace of the product space of four angular
momenta in which (\ref{Jsumvanishes}) holds as an operator equation.
We shall denote this subspace by $Z$.  According to
(\ref{Jsumvanishes}), the total angular momentum vanishes on $Z$.  The
orthonormality relations satisfied by the $6j$-symbol (see, for
example, Edmonds\cite{Edmonds60} Eq.~(6.2.9)) are essentially a
statement of the unitarity of the matrix $\langle j_{12} \vert j_{23}
\rangle$.

To be defined the $6j$-symbol (\ref{normal6j}) must satisfy four
triangle inequalities, in $(j_1,j_2,j_{12})$, $(j_2,j_3,j_{23})$,
$(j_3,j_4,j_{12})$, and $(j_1,j_4,j_{23})$, for example, $j_{12}$ must
lie between the bounds
	\begin{equation}
	|j_1-j_2| \le j_{12} \le j_1+j_2,
	\label{triangleinequal}
	\end{equation}
in integer steps.  For given $j_i$, $i=1,2,3,4$, these imply that
$j_{12}$ and $j_{23}$ vary between the limits
	\begin{eqnarray}
	j_{12,{\rm min}} &\le& j_{12} \le j_{12,{\rm max}}, 
	\nonumber \\
	j_{23,{\rm min}} &\le& j_{23} \le j_{23,{\rm max}},
	\label{j12j23inequals}
	\end{eqnarray}
in integer steps, where
	\begin{eqnarray}
	j_{12,{\rm min}} &=& \max (|j_1-j_2|,|j_3-j_4|),
	\nonumber \\
	j_{23,{\rm min}} &=& \max (|j_2-j_3|,|j_1-j_4|),
	\nonumber \\
	j_{12,{\rm max}} &=& \min (j_1+j_2,j_3+j_4),
	\nonumber \\
	j_{23,{\rm max}} &=& \min (j_2+j_3,j_1+j_4).
	\label{j12j23bounds}
	\end{eqnarray}
The number of allowed $j_{12}$ or $j_{23}$ values is the same, and it
is the dimension $D$ of the subspace $Z$ as well as the size of the
matrix $\langle j_{12} \vert j_{23} \rangle$,
	\begin{equation}
	D = \dim Z = j_{12,{\rm max}} - j_{12,{\rm min}} +1
                   = j_{23,{\rm max}} - j_{23,{\rm min}} +1.
	\label{dimZdef}
	\end{equation}

\subsection{Classical and Semiclassical Mechanics of the $6j$-symbol}
\label{cm6j}

The basic reference on the semiclassical mechanics of the $6j$-symbol
is Ponzano and Regge\cite{PonzanoRegge68}.  We add to their discussion
an appreciation of the Gram matrix (see
Appendix~\ref{constucttetrahedron}) and the recent realization that
the phase space of the $6j$-symbol is a sphere \cite{Charles08}.

We shall reserve lower case $j_i$ for quantum numbers as in
(\ref{normal6j}), and for semiclassical purposes we shall set
	\begin{equation}
	J_i = j_i+\frac{1}{2}
	\label{Jidef}
	\end{equation}
(with capital $J$'s), for $i=1,2,3,4,12,23$.  The quantity $J_i$ is
interpreted as the length of the classical angular momentum vector
associated with the quantum number $j_i$.  The correction $1/2$ is a
Maslov index, and the manner in which it arises in the semiclassical
theory of angular momentum is explained in Aquilanti et
al\cite{Aquilantietal07}.  

We shall write $\Jvec_i$ (in bold face) either for the vector of
angular momentum operators in quantum mechanics, or for the classical
angular momentum vector in a classical model.  The distinction will be
established by context.  For example, in the classical context,
(\ref{Jsumvanishes}) is interpreted as a constraint on the four
classical vectors $\Jvec_i$, $i=1,2,3,4$, while (\ref{J12J23defs}) is
interpreted as the definitions of two more classical angular momenta
$\Jvec_{12}$ and $\Jvec_{23}$.

The condition (\ref{Jidef}) is a quantization condition of the
Bohr-Sommerfeld type, restricting the the classical quantity $J_i$ to
discrete values.  Some authors have viewed (\ref{Jidef}) as an
approximation to $[j(j+1)]^{1/2}$, valid when $j$ is large, but in
fact it is represents the exact eigenvalues of a certain operator for
all values of $j$, even $j=0$.  When properly
understood\cite{Aquilantietal07}, (\ref{Jidef}) is equivalent to the
fact that the eigenvalues of the operator $J^2$ are $j(j+1)$.

In purely classical mechanics, however, there is no quantization, and
all variables take on continuous values.  We must allow this in order
to view the classical phase space.  To visualize the phase space of
the $6j$-symbol, we will assume that $J_i$, $i=1,2,3,4$ have any fixed
positive values, while $J_{12}$ and $J_{23}$ are variables, the
lengths of the vectors $\Jvec_{12}$ and $\Jvec_{23}$ defined by
(\ref{J12J23defs}).  Then $J_{12}$ and $J_{23}$ are restricted by
classical versions of the triangle inequalities,
	\begin{eqnarray}
	&J_{12,{\rm min}} \le J_{12} \le J_{12,{\rm max}}, 
	\nonumber \\
	&J_{23,{\rm min}} \le J_{23} \le J_{23,{\rm max}},
	\label{J12J23inequals}
	\end{eqnarray}
where $J_{12}$ and $J_{23}$ vary continuously between the bounds
indicated, and where the bounds themselves are given by
	\begin{eqnarray}
	J_{12,{\rm min}} &=& \max (|J_1-J_2|,|J_3-J_4|),
	\nonumber \\
	J_{23,{\rm min}} &=& \max (|J_2-J_3|,|J_1-J_4|),
	\nonumber \\
	J_{12,{\rm max}} &=& \min (J_1+J_2,J_3+J_4),
	\nonumber \\
	J_{23,{\rm max}} &=& \min (J_2+J_3,J_1+J_4).
	\label{J12J23bounds}
	\end{eqnarray}
These are the bounds of $J_{12}$ and $J_{23}$ on the $6j$-sphere.  

Here are two useful theorems.  First, if $J_{23,{\rm min}}=J_1-J_4$ or
$J_2-J_3$, then $J_{12,{\rm max}}=J_3+J_4$, otherwise $J_{12,{\rm
max}}=J_1+J_2$.  Second, if $J_{12,{\rm min}}=J_1-J_2$ or $J_4-J_3$,
then $J_{23,{\rm max}}=J_2+J_3$, otherwise $J_{23,{\rm max}}=
J_1+J_4$.

If the quantization conditions (\ref{Jidef}) hold for $J_i$,
$i=1,2,3,4$, then the bounds on the continuous variables $J_{12}$ and
$J_{23}$ can be expressed in terms of the bounds on the quantum number
$j_{12}$ and $j_{23}$ by
	\begin{eqnarray}
	&J_{12,{\rm min}} = j_{12,{\rm min}}, \qquad
	J_{12,{\rm max}} = j_{12,{\rm max}} +1, 
	\nonumber \\
	  &J_{23,{\rm min}} = j_{23,{\rm min}},
	 \qquad
	J_{23,{\rm max}} = j_{23,{\rm max}} +1.
	\label{Jjminmax}
	\end{eqnarray}
Combined with (\ref{dimZdef}), these imply
	\begin{equation}
	D=J_{12,{\rm max}}-J_{12,{\rm min}} = J_{23,{\rm max}}
	-J_{23,{\rm min}}.
	\label{Ddef}
	\end{equation}

If $J_i>0$, $i=1,2,3,4$ and if $J_{12}$ and $J_{23}$ satisfy the
triangle inequalities (\ref{J12J23inequals}), then it is possible to
find six classical vectors $\Jvec_i$, $i=1,2,3,4,12,23$ that can be
placed end-to-end in subsets of three to create four triangles.  The
triangles are defined by (\ref{J12J23defs}), plus $\Jvec_{12} +
\Jvec_3 + \Jvec_4=0$ and $\Jvec_1 + \Jvec_{23} + \Jvec_4=0$.  In 
particular, this can always be done when the six $J_i$ satisfy the
quantization conditions (\ref{Jidef}) for values of $j_i$ that are
valid in a $6j$-symbol, in which case $J_i>0$ for all six $i$ (because
$j_i\ge0$) and the areas of the triangles are positive (because the
$j_i$ satisfy the triangle inequalities).

For some values of the six $J_i$ the four triangles can be fitted
together to form the four faces of a tetrahedron.  This is the
classically allowed region of the $6j$-symbol.  If they can, then the
signed volume of the tetrahedron is given by
	\begin{equation}
	V = \frac{1}{6} \Jvec_1\cdot(\Jvec_2 \times \Jvec_3) =
	    \frac{1}{6} \Avec_1\cdot(\Avec_2 \times \Avec_3),
	\label{Vdef}
	\end{equation}
where the vectors $\Avec_i$ are defined in (\ref{Adefs}).
The volume is related to the nonnegative definite Gram
matrix $G$ defined in (\ref{Gijdef}) by
	\begin{equation}
	36 V^2 = \det G.
	\label{VdetG}
	\end{equation}
The tetrahedron is illustrated in Fig.~\ref{tetrahedron}, which shows
our convention for labeling the edges by classical angular momentum
vectors.  The vectors in the figure may be seen to satisfy
(\ref{Jsumvanishes}) and (\ref{J12J23defs}).  The tetrahedron in
Fig.~\ref{tetrahedron} has positive volume ($V>0$) according to the
definition (\ref{Vdef}).

For other values of the $J_i$ a real tetrahedron does not exist, but a
complex tetrahedron exists whose edges are complex vectors $\Jvec_i$.
These may be chosen so that the $x$- and $y$-components are purely
real and the $z$-component is purely imaginary.  This is the
classically forbidden region of the $6j$-symbol.  In this case $G$ is
still given by (\ref{Gijdef}) and it is still a real, symmetric
matrix, but it has one negative eigenvalue.  Also, $V^2<0$ and
$V$ is purely imaginary.  Equation~(\ref{Vdef}) is still valid,
however, in terms of the complex vectors $\Jvec_i$.  The dot product
of complex vectors $\Uvec$ and $\Vvec$ is defined by $\sum_i U_iV_i$
(not $\sum_i U_i^* V_i$), and the length by $J_i^2 = \Jvec_i
\cdot \Jvec_i$ (not $\Jvec_i^* \cdot \Jvec_i$).

The classically allowed and forbidden regions are illustrated in
Fig.~\ref{spots}.  This figure shows the square region of the
$J_{12}$--$J_{23}$ plane bounded by the classical limits
(\ref{J12J23bounds}) for certain fixed values of $j_i$, $i=1,2,3,4$.  The
small spots inside the square are the quantized values of $J_{12}$ and
$J_{23}$, from which the quantum numbers $j_{12}$ and $j_{23}$ can be
extracted by (\ref{Jidef}).  Notice that the quantized values of
$J_{12}$ and $J_{23}$ always stay at least one half unit away from the
bounds (\ref{J12J23bounds}).  The oval curve is the curve $V=0$,
separating the classically allowed from the classically forbidden
regions; it is the caustic curve.  The classically allowed and
forbidden regions lie inside and outside the caustic curve,
respectively.  Actually, there are four classically forbidden regions,
labeled ABCD in the figure.  Similar diagrams describe radiative
transitions in hydrogen (see Fig.~27 of Ref.~\cite{Braun93}).  

The caustic curve consists of real tetrahedra that are flat; it
touches the square at four points, labeled XYZW in Fig.~\ref{spots}.
The geometrical meaning of these points and the behavior of the flat
tetrahedron as we move around the caustic curve are illustrated in
Fig.~\ref{flat}.  At point Y, vectors $\Jvec_1$ and $\Jvec_2$ are
antiparallel, giving $\Jvec_{12}$ its minimum length.  As $\Jvec_2$
rotates in a counterclockwise direction, at first $\Jvec_{12}$ grows
and $\Jvec_{23}$ shrinks, as illustrated in the diagram
${\rm Y} \to {\rm X}$ in the figure.  This is a point between Y and X
on the caustic line of Fig.~\ref{spots}.  When $\Jvec_2$ rotates to
the angle that causes $\Jvec_2$ and $\Jvec_3$ to be antiparallel, then
$\Jvec_{23}$ is at its minimum length and we are at point X.
As $\Jvec_2$ continues to rotate, $\Jvec_{23}$ starts to grow
again while $\Jvec_{12}$ continues to grow, as illustrated in the
diagram ${\rm X} \to {\rm Z}$ in the figure.  This is a point between
X and Z on the caustic curve.  In this manner we may continue around
the caustic curve.

\subsection{The Ponzano-Regge Formula}
\label{PRformula}

Suppose we are in the classically allowed region so a real tetrahedron
exists.  Let $\psi_i$, $i=1,2,3,4,12,23$ be the exterior dihedral
angles of the tetrahedron associated with edge $\Jvec_i$, that is,
$\psi_i$ is the angle between the outward pointing normals of the two
faces that meet in edge $i$, so that each $\psi_i$ lies in the
interval $[0,\pi]$.  Then the Ponzano-Regge phase is defined
by
	\begin{equation}
	\Phi_{\rm PR} = \sum_i J_i \psi_i,
	\label{PhiPRdef}
	\end{equation}
where the sum runs over all six edges, and the Ponzano-Regge
approximation is
	\begin{equation}
	\left\{ \begin{array}{ccc}
	j_1 & j_2 & j_{12} \\
	j_3 & j_4 & j_{23}
	\end{array}\right\} \approx
	A_{\rm PR} \cos 
	\left(\Phi_{\rm PR} + \frac{\pi}{4}\right),
	\label{PRappx}
	\end{equation}
where the amplitude in the classically allowed region is given by
	\begin{equation}
	A_{\rm PR} = \frac{1}{\sqrt{12\pi |V|}}.
	\label{APRcar}
	\end{equation}

Given the six quantum numbers $j_i$, an algorithm for determining the
dihedral angles $\psi_i$ is the following.  The rules we give are
equivalent to those of Ponzano and Regge and Schulten and Gordon, but
stated in terms of the diagonalization of the Gram matrix.  First we
define the six $J_i$ by (\ref{Jidef}) and then we set up the Gram
matrix using (\ref{Gijcalc}) and diagonalize it.  If all the
eigenvalues are positive (or if $\det G = 36V^2 > 0$), we are in the
classically allowed region and we may proceed.  Then we construct the
six vectors $\Jvec_i$ as explained in
Appendix~\ref{constucttetrahedron}, we compute the outward pointing
normals by taking cross products of the vectors spanning the four
faces, and finally we compute $\cos\psi_i$ as the dot products of the
outward pointing normals.  This determines $\psi_i$ uniquely as an
angle in $[0,\pi]$.  This is not the most efficient algorithm from a
numerical standpoint, since to determine the $\psi_i$ only the dot
products of the vectors are needed and not the vectors themselves, but
it is conceptually clean and has the benefit of allowing one to draw
or visualize the tetrahedron itself.

The Ponzano-Regge phase $\Phi_{\rm PR}$ is continuous inside the
classically allowed region, as are the dihedral angles $\psi_i$.  On
the caustic boundary all tetrahedra are flat so all dihedral angles
are either 0 or $\pi$.  These angles are continuous (hence constant at
0 or $\pi$) on the caustic line between points XYZW, but some angles
jump discontinuously between 0 and $\pi$ at those points.  

The possible values of the $\psi_i$ on the caustic curve are
summarized in Table~\ref{psitable}.  The four segments of the caustic
curve are identified by the classically forbidden region (ABCD) to
which they are adjacent.  In segments A, B and D there are two
possibilities, while in segment C there is only one.  In segment A,
the first column applies if $J_{12,{\rm max}}=J_3+J_4$ and the second
column otherwise; in segment B, the first column applies if
$J_{23,{\rm min}} = J_2-J_3$ or $J_4-J_1$ and the second column
otherwise; and in segment D, the first column applies if
$J_{23,{\rm max}} = J_2+J_3$, and the second column otherwise.
Examples of these rules may be seen in Fig.~\ref{flat}.  The dihedral
angles in a flat tetrahedron such as the ones labeled ${\rm Y}\to{\rm
X}$ and ${\rm X}\to{\rm Z}$ are 0 for interior segments and $\pi$ for
segments bounding the outside of the plane figure.

\begin{table}
\begin{tabular}{|c||cc|cc|c|cc|}\hline
$i$ & A & A & B & B & C & D & D \\
\hline\hline
1 & $\pi$ & 0 & $\pi$ & 0 & $\pi$ & $\pi$ & 0 \\
2 & $\pi$ & 0 & 0 & $\pi$ & $\pi$ & 0 & $\pi$ \\
3 & $0$ & $\pi$ & $\pi$ & 0 & $\pi$ & 0 & $\pi$ \\
4 & $0$ & $\pi$ & 0 & $\pi$ & $\pi$ &$\pi$ & 0 \\
12 & $\pi$ & $\pi$ & $\pi$ & $\pi$ & 0 & 0 & 0 \\
23 & 0 & 0 & $\pi$ & $\pi$ & 0 & $\pi$ & $\pi$\\
\hline
\end{tabular}
\caption{\label{psitable} The dihedral angles $\psi_i$ on the 
segments of the caustic curve bounding classically forbidden regions 
ABCD.  There are two possibilities for segments B and D, and one for 
segments A and C.}
\end{table}

Although some angles $\psi_i$ are discontinuous at points XYZW, the
Pon\-zano-Regge phase $\Phi_{\rm PR}$ is continuous everywhere on the
caustic boundary (hence everywhere inside and on the caustic
boundary).  

The angles $\psi_i$ that are $\pi$ on a segment of the caustic curve
correspond to the vectors $\Jvec_i$ that lie on the outside of the
plane figure, as seen in the examples in Fig.~\ref{flat}.  That is,
they correspond to a set of vectors $\Jvec_i$ that sum to zero.  But
this implies that the sum of the corresponding $j_i$ values is an
integer,
	\begin{equation}
	\nu_{6j}={\mathop{\sum_i}\nolimits}' j_i = {\rm integer},
	\label{nudef}
	\end{equation}
where the prime means to sum only over $i$ such that $\psi_i$ on a
segment of the caustic curve has the value $\pi$.  For example, from
the first column for segment B in Table~\ref{psitable} we have
$j_1+j_3+j_{12}+j_{23} = {\rm integer}$.

In the classically forbidden regions the method of
Appendix~\ref{constucttetrahedron} yields vectors $\Jvec_i$ whose
$(x,y,z)$ components can be labeled as $(r,r,i)$, where $r$ means real
and $i$ means imaginary.  These vectors have real lengths $J_i$ that
are fixed by (\ref{Jidef}) and the values of the $j_i$.  The cross products
$\Jvec_i \times \Jvec_j$ have the form $(i,i,r)$ and also have
real lengths, which are twice the real areas of the faces.  Dividing by
these we obtain complex unit normals to the faces of the form
$(i,i,r)$, whose dot products, the cosines of the angles $\psi_i$, are
real.  These cosines lie outside the range $[-1,+1]$, however,
indicating that the $\psi_i$ are complex.  Since the complex inverse
cosine function has multiple branches, we must determine the branch.

A first requirement is that branch chosen for $\psi_i$ should agree
with the value of $\psi_i$ as we approach caustic curve, where
$\psi_i$ is either 0 or $\pi$, depending on $i$ and the region ABCD,
as shown in Table~\ref{psitable}.  If $\psi_i=0$, $\cos\psi_i=+1$ on
the caustic curve, then $\cos\psi_i$ is real and $>1$ in the
classically forbidden region.  In this case we choose
$\psi_i=i{\bar\psi}_i$, where ${\bar\psi}_i = \cosh^{-1} (\cos\psi_i)$
is real and positive.  If $\psi_i=\pi$, $\cos\psi_i=-1$ on the caustic
curve, then $\cos\psi_i$ is real and $<-1$ in the classically
forbidden region.  In this case we choose $\psi_i =
\pi+i{\bar\psi}_i$, where ${\bar\psi}_i = -\cosh^{-1}(-\cos\psi_i)$ is
real and negative.  We can summarize these two cases by
	\begin{equation}
	{\bar\psi}_i = \sign(\cos\psi_i)
	\cosh^{-1}(|\cos\psi_i|).
	\label{psibaridef}
	\end{equation}
In spite of the sign and absolute value functions, ${\bar\psi}_i$ is a
smooth function of position in any of the four classically forbidden
regions.

This procedure allows us to determine which classically forbidden
region (ABCD) we are in, for if we note the signs of the $\cos\psi_i$,
those that are $>1$ indicate angles that vanish on the segment of
the caustic curve bordering the region, while those that are $<-1$
indicate angles that become $\pi$ on that segment.  The pattern of
0's and $\pi$'s uniquely identifies the region, as shown by
Table~\ref{psitable}.

With these definitions, the imaginary part of the analytic
continuation of $\Phi_{\rm PR}$ is
	\begin{equation}
	{\bar\Phi}_{\rm PR}= \sum_i J_i {\bar \psi}_i,
	\label{barPhiPRdef}
	\end{equation}
where the sum runs over all six $i$.  The quantity ${\bar\Phi}_{\rm
PR}$ vanishes on the caustic curve and becomes real and negative as we
move into classically forbidden regions A or D, or real and positive
as we move into regions B or C.  The $6j$ symbol decays exponentially
as we move into any classically forbidden region, a behavior that is
captured by $\exp(-|{\bar\Phi}_{\rm PR}|)$ in all regions.  Since
$\Phi_{\rm PR}$ has only one sign in any of the four classically
forbidden regions, its absolute value is a smooth function in those
regions.  Finally, the Ponzano-Regge approximation in the classically
forbidden regions is
	\begin{equation}
	\left\{ \begin{array}{ccc}
	j_1 & j_2 & j_{12} \\
	j_3 & j_4 & j_{23}
	\end{array}\right\} \approx
	A_{\rm PR}
	\exp(-|{\bar\Phi}_{\rm PR}|), 
	\label{PRappxCFR}
	\end{equation}
where the amplitude is given by
	\begin{equation}
	A_{\rm PR} = \frac{(-1)^{\nu_{6j}}}{2\sqrt{12\pi|V|}},
	\label{APRcfr}
	\end{equation}
and where $\nu_{6j}$ is given by (\ref{nudef}).  The definitions we
have made allow a single formula to be written down for all four
regions, but it is easy to write four formulas for the four regions
without the use of sign or absolute value functions.

\subsection{The Phase Space of the $6j$-symbol}
\label{phasespace6j}

The phase space of the $6j$-symbol is the set of all closed figures
that can be obtained by placing $\Jvec_i$, $i=1,2,3,4$ end-to-end, for
fixed values of the lengths of these vectors, modulo proper rotations.
That is, the vectors must satisfy (\ref{Jsumvanishes}).  This is the
point of view of Kapovich and Millson\cite{KapovichMillson96}, that
recently has been further developed by Charles\cite{Charles08}.  This
space can also be derived by symplectic
reduction\cite{AbrahamMarsden78} from a model of four independent
angular momenta built around Schwinger's\cite{BiedenharnvanDam65}
oscillators, much as in the treatment of Aquilanti et
al\cite{Aquilantietal07} of the $3j$-symbol.  In our applications we
will think of the lengths $J_i$, $i=1,2,3,4$ as being given by the
quantization condition (\ref{Jidef}) in terms of the fixed quantum
numbers $j_i$, $i=1,2,3,4$ appearing in a $6j$-symbol.  For a given
closed chain formed by $\Jvec_i$, $i=1,2,3,4$, we can draw vectors
$\Jvec_{12}$ and $\Jvec_{23}$ defined by (\ref{J12J23defs}) to obtain
a tetrahedron.  Thus the phase space can also be thought of as the set
of all real tetrahedra, in which the lengths $J_i$, $i=1,2,3,4$ are
fixed.  The lengths $J_{12}$ and $J_{23}$, however, vary continuously
between the limits (\ref{J12J23bounds}).

All such tetrahedra are generated if we let $J_{12}$ vary from
$J_{12,{\rm min}}$ to $J_{12,{\rm max}}$, while for each value of
$J_{12}$ we let the dihedral angle $\phi_{12}$ vary from $-\pi$ to
$+\pi$.  Here $\phi_{12}$ is the interior dihedral angle,
illustrated in Fig.~\ref{dihedral}, that is uniquely defined in the
interval $-\pi < \phi_{12} \le \pi$ by requiring that $\phi_{12}=0,\pi$
correspond to flat tetrahedra, and that $0< \phi_{12}<\pi$ correspond to
tetrahedra with positive volume (this is the case illustrated in
Fig.~\ref{dihedral}).  It is related to the exterior dihedral angle
$\psi_{12}$ used in the Ponzano-Regge formula by
$|\phi_{12}|+\psi_{12} = \pi$.  The angle $\phi_{12}$
distinguishes tetrahedra related by spatial inversion (i.e., time
reversal), while $\psi_{12}$ does not.  Similarly, we could generate
all real tetrahedra by varying $J_{23}$ and the dihedral angle
$\phi_{23}$.  The choice of $J_{12}$, $\phi_{12}$ for this process
is arbitrary, but it gives us a definite convention for coordinates on
the phase space of the $6j$-symbol, namely, $(J_{12},\phi_{12})$.

The manifold of such tetrahedra, modulo proper rotations, is a
sphere.  To see this, define
	\begin{equation}
	J_{12,{\rm avg}} = \frac{1}{2}(J_{12,{\rm max}} + J_{12,{\rm
	min}}),
	\label{J12avgdef}
	\end{equation}
and write
	\begin{equation}
	K_z = J_{12} - J_{12,{\rm avg}},
	\label{Kzdef}
	\end{equation}
so that $K_z$ varies between $-D/2$ and $+D/2$ as $J_{12}$ goes from
$J_{12,{\rm min}}$ to $J_{12,{\rm max}}$ (see (\ref{Ddef})).  Then 
define a polar angle $\theta_{12}$ by
	\begin{equation}
	K_z=(D/2)\cos\theta_{12},
	\label{Kperpdef}
	\end{equation}
and set
	\begin{eqnarray}
	K_x &=& (D/2)\sin\theta_{12}\cos\phi_{12}, \nonumber \\
	K_y &=& (D/2)\sin\theta_{12}\sin\phi_{12},
	\label{KxKydefs}
	\end{eqnarray}
so that $(K_x,K_y,K_z)$ are Cartesian coordinates on a sphere of
radius $D/2$ with spherical angles $(\theta_{12},\phi_{12})$.  The
azimuthal angle $\phi_{12}$ on the sphere is the same as the interior
dihedral angle in the tetrahedron.

This is the $6j$-sphere, on which the north pole is $K_z=D/2$ or
$J_{12} = J_{12,{\rm max}}$, the south pole is $K_z=-D/2$ or $J_{12} =
J_{12,{\rm min}}$, and curves of constant $J_{12}$ in general are
small circles $K_z ={\rm const}$.  It is illustrated in
Fig.~\ref{6jsphere}, which shows several curves of constant $J_{12}$.
Flat configurations correspond to $\phi_{12}=0$ or $\pi$, that is,
they lie on the $K_x$--$K_z$ plane (the great circle $K_y=0$).  The
hemisphere $K_y>0$ ($K_y<0$) consists of tetrahedra of positive
(negative) volume.  Spatial inversion (i.e., time reversal) is a reflection
in the plane $K_y=0$ (it amounts to $K_y
\to -K_y$).

Any quantity defined in terms of the tetrahedron that is invariant
under proper rotations corresponds to a function on the $6j$-sphere.
For example, $J_{23}$ is such a function, as is $\phi_{23}$.  Curves
of constant $J_{23}$ are illustrated in Fig.~\ref{J23orbits}.  The
extrema of $J_{23}$ are both flat configurations lying on the great
circle $K_y=0$, with $J_{23}=J_{23,{\rm min}}$ on the semicircle
$K_x>0$ and $J_{23}=J_{23,{\rm max}}$ on the semicircle $K_x<0$.  This
is apparent from figures such as diagram X in Fig.~\ref{flat}, which
illustrates the case $J_{23}=J_{23,{\rm min}}$ and which shows that
$\phi_{12}=0$ (not $\pi$) at such a configuration.  Two views of the
$6j$-sphere are given in Fig.~\ref{J23orbits} to show both extrema of
$J_{23}$, as well as curves of constant $J_{23}$ for intermediate
values.  Notice that the curves of constant $J_{23}$ are not small
circles.  That is because the coordinates we are using on the
$6j$-sphere make the curves of constant $J_{12}$ look simple (they are
small circles), but not the curves of constant $J_{23}$.  Had we based
our coordinates on $J_{23}$ and $\phi_{23}$ instead, the roles would
be reversed.

The diagrams in Fig.~\ref{J23orbits} were created in the following
way.  We set up a grid of coordinates $(J_{12},\phi_{12})$ by
letting $J_{12}$ vary between $J_{12,{\rm min}}$ and $J_{12,{\rm
max}}$, and for each value of $J_{12}$, letting $\phi_{12}$ vary
between $\pm\pi$.  For each value of $J_{12}$ at $\phi_{12}=0$, we
set up the corresponding flat tetrahedron, such as the diagrams
labelled ${\rm Y}\to{\rm X}$ and ${\rm X}\to{\rm Z}$ in
Fig.~\ref{flat}, both of which have $\phi_{12}=0$.  For other values
of $\phi_{12}$, we rotate the triangle 1--2--12 about the axis
defined by $\Jvec_{12}$ by angle $\phi_{12}$, using the right-hand
rule, while holding triangle 3--4--12 fixed.  This is the
``butterfly'' motion of the tetrahedron associated with the axis
$\Jvec_{12}$.  This creates a tetrahedron of the desired dihedral
angle $\phi_{12}$.  Then we compute $J_{23}$ for that tetrahedron.
In this way, we set up an array of $J_{23}$ values on the grid.
Finally, we draw the contour lines of $J_{23}$ for this grid, and plot
them on the surface of the sphere.

There are two ways to compute Poisson brackets on the $6j$-sphere.
First, let $F$ and $G$ be two functions of the $\Jvec_i$, $i=1,2,3,4$.
Then the Poisson bracket is the usual one in classical mechanics for a
set of independent angular momenta,
	\begin{equation}
	\{F,G\} = \sum_{i=1}^4 \Jvec_i \cdot
	\left( \frac{\partial F}{\partial \Jvec_i} \times
	\frac{\partial G}{\partial \Jvec_i} \right).
	\label{PBdef}
	\end{equation}
For example, any function of $\Jvec_1$ and $\Jvec_2$ has vanishing
Poisson bracket with any function of $\Jvec_3$ and $\Jvec_4$.

Hamilton's equations can be expressed in terms of Poisson brackets.
Let $H$ be a Hamiltonian with evolution parameter (the ``time'')
$\lambda$, and let $F$ be any function of $\Jvec_i$, $i=1,2,3,4$.
Then the rate of change of $F$ along the orbits of $H$ is
	\begin{equation}
	\frac{dF}{d\lambda} = \{F,H\}.
	\label{Hameqns}
	\end{equation}
For example, if we take $H=J_{12} = |\Jvec_1+\Jvec_2|$ and $F=$ some
component of one of the $\Jvec_i$, we find
	\begin{equation}
	\frac{d\Jvec_i}{d\lambda} = \left\{
	\begin{array}{r@{\quad}l}
	\evec_{12}\times \Jvec_i, & i=1,2, \\
	0, & i=3,4,
	\end{array}
	\right.
	\label{bflyodes}
	\end{equation}
where $\evec_{12}$ is the unit vector in the direction $\Jvec_{12}$.
The $\lambda$-evolution is a rotation of vectors $\Jvec_1$ and
$\Jvec_2$ about the axis $\evec_{12}$, while $\Jvec_3$ and $\Jvec_4$
remain fixed.  This is the ``butterfly'' motion mentioned above, and
$\lambda$ is the angle.  If the initial conditions are chosen so that
$\lambda=0$ when $\phi_{12}=0$, then $\lambda=\phi_{12}$.  On the
$6j$-sphere the orbits of $J_{12}$ are the curves $J_{12}={\rm
const}$, the small circles seen in Fig.~\ref{6jsphere}.

Similarly, $J_{23}$ generates another butterfly motion, in which
vectors $\Jvec_2$ and $\Jvec_3$ rotate about the axis $\evec_{23}$ by
the right-hand rule, with dihedral angle $\phi_{23}$ as the parameter
of evolution.  Vectors $\Jvec_1$ and $\Jvec_4$ remain fixed during
this motion.  The orbits on the $6j$-sphere are curves of constant
$J_{23}$, some examples of which are illustrated in
Fig.~\ref{J23orbits}.

Since a Hamiltonian and its evolution parameter are canonically
conjugate variables, it follows that $(J_{12},\phi_{12})$ are
canonically conjugate variables on the $6j$-sphere.  Since $J_{12}$
differs from $K_z$ by a constant, we can equally well use
$(K_z,\phi_{12})$.  Thus another way to compute the Poisson bracket
of any two functions defined on the $6j$-sphere is
	\begin{equation}
	\{F,G\} = \frac{\partial F}{\partial\phi_{12}}
	\frac{\partial G}{\partial J_{12}}
	-\frac{\partial F}{\partial J_{12}}
	\frac{\partial G}{\partial \phi_{12}}.
	\label{altPBdef}
	\end{equation}
This only applies to rotationally invariant functions, while
(\ref{PBdef}) can be used for any functions of the $\Jvec_i$,
$i=1,2,3,4$.  Also, expressing functions in terms of $J_{12}$ and
$\phi_{12}$ is usually difficult, so in practice (\ref{PBdef}) is
more useful than (\ref{altPBdef}).  

But (\ref{altPBdef}) does show that the area of a closed curve on the
$6j$-sphere can be computed as
	\begin{equation}
	{\rm Area} = \oint J_{12}\,d\phi_{12},
	\label{6jarea}
	\end{equation}
with due attention to the singularities of the $(J_{12},\phi_{12})$
coordinates (there is no global symplectic 1-form on the sphere).  The
area is $D/2$ times the solid angle subtended by the closed loop.
This is not the area on the surface of a sphere of radius $D/2$ in
$(K_x,K_y,K_z)$ space, computed by Euclidean geometry, which would be
$(D/2)^2$ times the solid angle, but it is the correct measure of area
from the standpoint of semiclassical mechanics.  

In particular, the total area of the sphere is $(4\pi)(D/2) = 2\pi D$,
or $D$ Planck cells of area $2\pi$ each (it would be $2\pi\hbar$ in
ordinary units).  This is what we expect for a semiclassical phase
space representing the subspace $Z$ of the Hilbert space of four
angular momenta, which contains $D$ quantum states.

Moreover, the Bohr-Sommerfeld rules say that the quantized values of
$J_{12}$ should be given by orbits $J_{12}={\rm const}$ whose area is
$(n+\frac{1}{2})(2\pi)$.  On the sphere there is no way to distinguish
the interior and the exterior of a loop, but the Bohr-Sommerfeld rule
is the same in either case since the total area of the sphere is an
integer times $2\pi$.  Since the small circle $J_{12}={\rm const}$
encloses area $(2\pi)(J_{12}-J_{12,{\rm min}})$ about the south pole,
the quantized orbits are those for which $J_{12} = J_{12,{\rm
min}}+\frac{1}{2} +{\rm integer}$.  By (\ref{Jidef}) this gives the
exact quantized values of $j_{12}$, as indicated by
(\ref{triangleinequal}).  The minimum and maximum quantized values of
$J_{12}$ are one half unit away from the values at the south and north
poles, respectively, corresponding to the one-half unit margin between
the quantized spots in Fig.~\ref{spots} and the bounding values of
$J_{12}$.  Figure~\ref{6jsphere} illustrates the quantized orbits of
$J_{12}$ for the case $D=5$, numbered 0 to 4 as $J_{12}$ increases
(orbit 0 lies close to the south pole and cannot be seen in the
figure).

Similarly, the quantized orbits of $J_{23}$ are those satisfying
$J_{23} = J_{23,{\rm min}}+\frac{1}{2}+{\rm integer}$.  These are
labeled 0 to 4 in Fig.~\ref{J23orbits}, for the same case ($D=5$)
illustrated in Fig.~\ref{6jsphere}.  The dihedral angle $\phi_{23}$
is an angle parameterizing position along the curves $J_{23}={\rm
const}$, although it is not an azimuthal angle in $(K_x,K_y,K_z)$
space.  

But the logic that leads to the conclusion that $(J_{12},\phi_{12})$
are canonical coordinates on the sphere applies also to
$(J_{23},\phi_{23})$, so there is a canonical transformation
connecting $(J_{12},\phi_{12})$ and $(J_{23},\phi_{23})$.  According
to Miller's theory\cite{Miller74}, the $F_4$-type generating function
of this canonical transformation is the phase of the semiclassical
matrix element $\langle j_{12} \vert j_{23} \rangle$.  (We follow
Goldstein's\cite{Goldstein80} conventions for classifying generating
functions.)

Miller's theory leads to difficult integrals in cases like this (it
certainly does for the $3j$-symbol), and it has never been carried
through for the $6j$-symbol, as far as we know.  But it is certain
that the $F_4$-type generating function that would result would be the
Ponzano-Regge phase $\Phi_{\rm PR}$, to within an additive constant.
Moreover, Miller's theory shows that the amplitude determinant in the
Ponzano-Regge formula is given by
	\begin{equation}
	A_{\rm PR} = \left| \frac{\partial^2 \Phi_{\rm PR}} 
	{\partial J_{12}\partial J_{23}}\right|^{1/2},  
	\label{PRamplitude} 
	\end{equation}
to within a multiplicative constant.  In this formula, the lengths
$J_i$, $i=1,2,3,4$ are considered fixed, and the dihedral angles
$\psi_i$ that appear in (\ref{PhiPRdef}) are considered functions of
all six lengths $J_i$.

The Ponzano-Regge amplitude was first derived by
Wigner\cite{Wigner59}, who had the intuition that the probability in
making a measurement of $j_{23}$ for given value of $j_{12}$ should be
uniformly distributed in the angle $\phi_{12}$.  That this is so
follows from standard semiclassical theory and the fact that
$\phi_{12}$ is conjugate to $J_{12}$.  This amplitude is also
inversely proportional to the square root of a Poisson bracket,
	\begin{equation} 
	\{J_{23},J_{12}\} = \frac{\Jvec_1\cdot(\Jvec_2\times\Jvec_3)}
	{J_{12}J_{23}} = \frac{6V}{J_{12}J_{23}},
	\label{6jamplitudePB}
	\end{equation}
which was computed using (\ref{PBdef}).  The volume factor $V$ is the
part of the amplitude that was obtained by Wigner, while the factor
$J_{12}J_{23}$, when replaced by $(j_{12}+\frac{1}{2})
(j_{23}+\frac{1}{2})$, is needed to convert from the unitary matrix
element $\langle j_{12} \vert j_{23} \rangle$ to the $6j$-symbol (see
Edmonds\cite{Edmonds60} Eq.~(6.2.10)).  The use of Poisson brackets
for computing amplitude determinants is discussed in
Littlejohn\cite{Littlejohn90} and in Aquilanti et
al\cite{Aquilantietal07}.

The relation between Fig.~\ref{spots} and the phase space of the 
$6j$-symbol is the following.  A point inside the square of
Fig.~\ref{spots} specifies values of $J_{12}$ and $J_{23}$ that are
allowed by the inequalities (\ref{J12J23inequals}).   These in turn
specify two curves in the phase space, one of constant $J_{12}$ and
the other of constant $J_{23}$.  If these curves intersect, as in part
(a) of Fig.~\ref{carcfr}, then we are in the classically allowed
region.  In that case, the intersection points, labeled $P$ and $Q$ in
the figure, are the stationary phase points of the semiclassical
evaluation of the matrix element $\langle j_{12} \vert j_{23}
\rangle$.  These points represent two tetrahedra that are mirror
images of each other (they are related by time reversal).  The total
semiclassical matrix element (the Ponzano-Regge formula) is a sum of
contributions from these two tetrahedra, which are complex conjugates
of each other.  Thus, the semiclassical matrix element is real.

If the two curves do not intersect, as in part (b) of
Fig.~\ref{carcfr}, then we are in the classically forbidden region.
Both curves are manifolds of real tetrahedra, one with a fixed value
of $J_{12}$, the other of $J_{23}$, but since they do not intersect,
there is no real tetrahedron that simultaneously has both given values
of $J_{12}$ and $J_{23}$.  In this case the analytic continuations of
the curves into complex phase space (a complexified sphere) do
intersect, and these intersections represent the stationary phase
points in the classically forbidden region.  We make no attempt to
sketch the complexified phase space, however.

In addition to its interpretation as a generating function, the phase
of semiclassical matrix elements such as $\langle j_{12} \vert j_{23}
\rangle$ is geometrically one half the area enclosed by the intersection of the
quantized classical orbits in phase space\cite{Littlejohn90}.  (The
relative phase between the two branches of the WKB solution is the
area, but this is shared between two exponentials to create a cosine
term.  Thus the argument of the cosine is one half the area.)  In part
(a) of Fig.~\ref{carcfr} the area in question is the shaded region (a
``lune''). 

If the two curves $J_{12}={\rm const}$ and $J_{23}={\rm const}$ are
tangent, then we are at a caustic.  A caustic implies a flat
tetrahedron of zero volume, so such tangencies can occur only in the
plane $K_y=0$.  

The different types of caustics than can occur are illustrated in
Fig.~\ref{caustics}.  Part~A of Fig.~\ref{caustics} is obtained from
the curves of Fig.~\ref{carcfr} by adjusting the $J_{12}$ or $J_{23}$
values to create a tangency.  As expected, it lies between the
classically allowed region (part~(a) of Fig.~\ref{carcfr}) and the
classically forbidden region (part~(b) of Fig.~\ref{carcfr}).  The
point of tangency $T$ is the caustic point.  From part~A of
Fig.~\ref{caustics} we move into the classically allowed region if
we either decrease $J_{12}$ or increase $J_{23}$.  Thus we see that it
corresponds to region~A of Fig.~\ref{spots}.

If we allow $J_{12}$ in part~A of Fig.~\ref{caustics} to decrease, the
small circles $J_{12}={\rm const}$ sweep down from the north pole
through the oval $J_{23}={\rm const}$, passing through the classically
allowed region, until a tangency is reached at the lower point of the
oval $J_{23}={\rm const}$, where the small circle $J_{12}={\rm const}$
is close to the equator.  This is another caustic, illustrated in
part~B of Fig.~\ref{caustics}.  The shaded area is the continuation of
the shaded area in part~(a) of Fig.~\ref{carcfr}, and again $T$ is the
caustic point.  From this caustic, we pass back into the classically
allowed region if either $J_{12}$ increases or $J_{23}$ increases, so
we are in region~B of Fig.~\ref{spots}.

If from part~A of Fig.~\ref{caustics} we allow $J_{23}$ to increase,
then the curves $J_{23}={\rm const}$ sweep through the small circle
$J_{12}={\rm const}$ about the north pole, finally reaching another
tangency on the other side where $K_x<0$.  The result is illustrated
in part~C of Fig.~\ref{caustics}, where again point $T$ is the caustic
point.  From this caustic we pass back into the classically allowed
region if we let either $J_{12}$ or $J_{23}$ decrease, so this
corresponds to region~C of Fig.~\ref{spots}.  

As $J_{23}$ increases from its value in part~C of Fig.~\ref{caustics},
the curve $J_{23}={\rm const}$ shrinks down around the point
$J_{23}=J_{23,{\rm max}}$ on the semicircle $K_y=0$, $K_x<0$.  Then
allowing $J_{12}$ to decrease, the small circle around the north pole
moves south, passing through the curve $J_{23}={\rm const}$, producing
finally a tangency $T$ on the other side, as illustrated in part~D of
Fig.~\ref{caustics}.  Now the shaded area (twice the Ponzano-Regge phase
plus a constant) covers nearly the entire sphere.  From this
configuration we pass back into a classically allowed region if either
$J_{23}$ decreases or $J_{12}$ increases, so we are in region~D in
Fig.~\ref{spots}.

In the case of an ordinary oscillator with a flat phase space (the
plane), the difference in the actions between the two turning points
is one half the area of the orbit, and has the form
$(n+\frac{1}{2})\pi$, where $n$ is an integer.  As explained in the
introduction, this is a requirement for the existence of a uniform
approximation of the Weber function type.  The analogous statement for
the $6j$-symbol with the spherical phase space is sometimes true, and
sometimes not.  A case where it is true is obtained from diagrams A
and B of Fig.~\ref{caustics} in which we regard $j_{23}$ fixed and
$j_{12}$ variable.  As $J_{12}$ decreases from the north pole (its
maximum value), we first encounter a caustic of the type A, where the
area of the lune is zero.  Continuing to decrease $J_{12}$, we pass
through the classically allowed region, finally encountering a caustic
of the type B (the lower turning point), where the area of the lune is
the quantized area of the oval $j_{23}={\rm const}$.  This area has
the form $(n+\frac{1}{2})2\pi$, so the differences in the actions at
the two turning points is quantized.  This implies that the difference
in the Ponzano-Regge phases $\Phi_{\rm PR}$ between the two turning
points is also quantized.  

A case where the differences in the actions is not quantized and a
uniform approximation of the Weber function type does not exist is
obtained when $j_{23}$ has a value such as that illustrated in part C
of Fig.~\ref{caustics}.  In this case, as we let $J_{12}$ decrease
from its maximum value at the north pole the first caustic we
encounter is of type C, where the area of the lune is the area of the
curve $J_{12}={\rm const}$ (the shaded area in part C of the figure).
This area is not quantized, since the value of $J_{12}$ at a caustic
is not quantized.  As $J_{12}$ decreases, we eventually reach the
lower caustic of type B, where the area of the lune is the quantized
area of the orbit $j_{23}={\rm const}$.  Thus, the differences between
the areas, one quantized, the other not, is not quantized.  A case
like this (with caustics of the type B and C) was illustrated in
Fig.~\ref{noweber}.  

We will now show that the $d$-matrices have a phase space and an orbit
and caustic structure that are identical, from a topological
standpoint, to those of the $6j$-symbol.

\section{The $d$-matrices}
\label{dmatrices}
\subsection{Quantum Mechanics of the $d$-Matrices}

The $d$-matrices are defined by
	\begin{equation}
	d^j_{mm'}(\beta) = \langle m \vert U_y(\beta) \vert m'\rangle,
	\label{djmmdef}
	\end{equation}
where $U_y(\beta)=\exp(-i\beta J_y)$ is a rotation operator with Euler angle
$\beta$ about the $y$-axis, and $\vert m\rangle$ and $\vert m'\rangle$
are standard rotation basis states (eigenstates of $J_z$).  To
indicate both the operator and the quantum number, we will write these
states as $\vert J_z: m\rangle$ and $\vert J_z: m'\rangle$.  By
conjugation the rotation operator $U_y(\beta)$ rotates the angular
momentum vector, 
	\begin{equation}
	U_y(\beta)^\dagger \Jvec U_y(\beta) = R_y(\beta)\Jvec,
	\label{adjointeqn}
	\end{equation}
where $R_y(\beta)$ is the $3\times3$ rotation matrix for an active
rotation about the $y$-axis.  We define
	\begin{equation}
	\nhat = R_y(\beta)\zhat = 
	\left(\begin{array}{ccc}
	\cos\beta & 0 & \sin\beta \\
	0 & 1 & 0 \\
	-\sin\beta & 0 & \cos\beta
	\end{array}\right)
	\left(\begin{array}{c}
	0 \\ 0 \\ 1
	\end{array}\right)=
	\left( \begin{array}{c}
	\cos\beta \\
	0 \\
	\sin\beta
	\end{array}\right),
	\label{nhatdef}
	\end{equation}
as illustrated in Fig.~\ref{euler}, so that
	\begin{equation}
	(\nhat\cdot\Jvec) U_y(\beta) \vert J_z: m'\rangle
	= U_y(\beta) (\zhat\cdot\Jvec)\vert J_z: m'\rangle
	= m' U_y(\beta) \vert J_z: m'\rangle,
	\label{rotateket}
	\end{equation}
where we use (\ref{adjointeqn}) and
	\begin{equation}
	\nhat\cdot[R_y(\beta)\Jvec] = [R_y(\beta)^{-1}\nhat]\cdot\Jvec
	= \zhat\cdot\Jvec = J_z.
	\end{equation}
Therefore $U_y(\beta)\vert J_z:m'\rangle$ is an eigenstate of
$\nhat\cdot\Jvec \equiv J_n$ with eigenvalue $m'$, and we will write
	\begin{equation}
	U_y(\beta)\vert J_z: m'\rangle = \vert J_n: m'\rangle,
	\label{Jneigenstate}
	\end{equation}
so that
	\begin{equation}
	d^j_{mm'}(\beta) = \langle J_z: m \vert J_n: m'\rangle.
	\label{djmmme}
	\end{equation}
In this way the $d$-matrix is written as a unitary matrix element
connecting the eigenstates of two different operators.  This is the
starting point for Miller's\cite{Miller74} theory of semiclassical
matrix elements, as well as our own\cite{Littlejohn90,Aquilantietal07}
treatments of the same subject.

\subsection{Classical and Semiclassical Mechanics of the $d$-ma\-tri\-ces}
\label{SCMdmatrices}

References on the semiclassical approximation for the $d$-matrices
include Brussaard and Tolhoek\cite{BrussardTolhoek57}, Ponzano and
Regge\cite{PonzanoRegge68}, Braun et al\cite{Braunetal96} and
So\-ko\-lov\-ski and Connor\cite{SokolovskiConnor99}.  In the following we
emphasize geometrical aspects of the problem not covered by these
authors.  

The classical phase space for $d^j_{mm'}(\beta)$ is a sphere (``the
$d$-sphere'') in angular momentum space of radius $|\Jvec|=J$, where
	\begin{equation}
	J=j+\frac{1}{2}.  
	\label{Jdef}
	\end{equation}
The area of a loop on the surface of the sphere is given by
	\begin{equation}
	{\rm Area} = \oint J_z\, d\phi,
	\label{dArea}
	\end{equation}
where $\phi$ is the azimuthal angle, again with due consideration of
the singularities of the coordinates $(\phi,J_z)$.  That is, if the loop
subtends solid angle $\Omega$, then the area is $J\Omega$ (not
$J^2\Omega$, as in Euclidean geometry).  The total area of the sphere
is therefore $4\pi J = (2j+1)(2\pi)$, that is, the sphere consists of
$2j+1$ Planck cells, corresponding to the $2j+1$ basis states $\vert
J_z:m\rangle$ or $\vert J_n: m'\rangle$.  Curves of constant
$\zhat\cdot \Jvec=J_z$ and $\nhat\cdot\Jvec = J_n$ are small circles
centered on the axes $\zhat$ and $\nhat$, respectively, as illustrated
in Fig.~\ref{znorbits}.

The Poisson bracket of two functions $F$ and $G$ of $\Jvec$ is
(\ref{PBdef}) with a single term in the sum,
	\begin{equation}
	\{F,G\} = \Jvec\cdot\left(
	\frac{\partial F}{\partial\Jvec} \times
	\frac{\partial G}{\partial\Jvec}\right),
	\label{dPBdef}
	\end{equation}
or, equivalently, for functions of $(\phi,J_z)$,
	\begin{equation}
	\{F,G\} = \frac{\partial F}{\partial \phi}
	\frac{\partial G}{\partial J_z} -
	\frac{\partial F}{\partial J_z}
	\frac{\partial G}{\partial \phi}.
	\label{altdPBdef}
	\end{equation}
For example, using (\ref{dPBdef}) we find that Hamilton's equations for
Hamiltonian $J_z=\zhat\cdot\Jvec$ with evolution parameter $\lambda$
are	
	\begin{equation}
	\frac{d\Jvec}{d\lambda} = \zhat \times \Jvec.
	\label{Jzeqn}
	\end{equation}
The motion is a rotation about the $z$-axis, so the orbits are the
small circles $J_z={\rm const}$, as expected.  The parameter of the
orbit is $\lambda=\phi$, so $\phi$ and $J_z$ are conjugate variables,
as indicated in (\ref{altdPBdef}).  Similarly, $J_n$ generates
rotations about the axis $\nhat$.

Let $(\theta,\phi)$ be the usual spherical coordinates referred to the
axis $\zhat$, and let $(\theta',\phi')$ be an alternative set referred
to the axis $\nhat$.  That is, the $(\theta',\phi')$ coordinates of a
point $(x,y,z)$ are the same as the $(\theta,\phi)$ coordinates of the
inverse rotated point $R_y(\beta)^{-1}(x,y,z)$.  Thus the coordinate
transformation $(\theta,\phi) \to (\theta',\phi')$ is specified by
	\begin{eqnarray}
	\sin\theta \cos\phi &=& \cos\beta\sin\theta'\cos\phi'
	+\sin\beta\cos\theta', \nonumber \\
	\sin\theta \sin\phi &=& \sin\theta'\sin\phi',
	\nonumber \\
	\cos\theta &=& -\sin\beta\sin\theta'\cos\phi' +
	\cos\beta\cos\theta'.
	\label{thetaphixfm}
	\end{eqnarray}
The azimuthal angle $\phi'$ is conjugate to $J_n$, so both
$(\phi,J_z)$ and $(\phi',J_n)$ are canonical coordinates on the
sphere.  The $F_4$-type generating function of the canonical
transformation between these coordinates is the phase of the
semiclassical approximation to the $d$-matrices, according to Miller's
theory.  This aspect of the problem has been developed by
So\-ko\-lov\-ski and Connor\cite{SokolovskiConnor99}.

The classical observables $J_z$ and $J_n$ are functions on the
$d$-sphere that vary continuously between the limits,
	\begin{equation}
	-J \le J_z,J_n \le +J.
	\label{JzJnbounds}
	\end{equation}
The quantized orbits of $J_z$ and $J_n$ are those enclosing
$n+\frac{1}{2}$ Planck cells where $n$ is an integer. This implies
$J_z=m$ and $J_n=m'$ with the usual rules for quantum numbers $m$ and
$m'$,
	\begin{equation}
	-j \le m, m' \le +j,
	\label{mmprimebounds}
	\end{equation}
in integer steps.  Thus the maximum and minimum values of $m$ and $m'$
lie one half unit away from the maximum and minimum values of the
classical observables $J_z$ and $J_n$, as illustrated in
Fig.~\ref{dspots}.  This figure may be compared to Fig.~\ref{spots}
for the $6j$-symbol.  See also Fig.~1 of Braun et al\cite{Braunetal96}. 

When the $J_z$-orbit and the $J_n$-orbit intersect one another as in
part (a) of Fig.~\ref{znorbits}, then we are in the classically
allowed region of the $d$-matrices.  There are generically two
intersection points related by a reflection in the plane $J_y=0$,
marked by unit vectors from the origin $\ahat$ and $\ahat'$ in
Fig.~\ref{intersect}.  We concentrate on intersection $\ahat$, for
which $J_y>0$; at the other intersection $\ahat'$ we have $J_y<0$.  In the
coordinate systems $(\theta,\phi)$, $(\theta',\phi')$, the $\theta$
and $\theta'$ coordinates of intersection $\ahat$ are given by
	\begin{equation}
	\cos\theta = \frac{J_z}{J} = \frac{m}{j+\frac{1}{2}},
	\qquad
	\cos\theta' = \frac{J_n}{J} = \frac{m'}{j+\frac{1}{2}},
	\label{thetathetaprimedef}
	\end{equation}
where $J_z$ and $J_n$ label the two small circles and where the second
form applies if $J$, $J_z$ and $J_n$ take on their quantized values.
As for the $\phi$ and $\phi'$ coordinates of intersection point
$\ahat$, they can be obtained by solving (\ref{thetaphixfm}), assuming
$\theta$, $\theta'$ and $\beta$ are given.  This gives
	\begin{equation}
	\cos\phi=\frac{\cos\theta'-\cos\beta\cos\theta}
	{\sin\beta \sin\theta}, 
	\qquad
	\cos\eta = \frac{\cos\theta-\cos\beta \cos\theta'}
	{\sin\beta \sin\theta'},
	\label{phiphiprimedef}
	\end{equation}
where we write $\eta=\pi-\phi'$ as illustrated in
Fig.~\ref{intersect}.  Equations~(\ref{phiphiprimedef}) uniquely
determine $\phi$ and $\eta$ in the interval $[0,\pi]$ (assuming that
the $J_z$- and $J_n$-orbits actually intersect).

Figure~\ref{intersect} draws attention to the spherical triangle
defined by $\zhat$, $\nhat$ and $\ahat$, whose sides are arcs of great
circles subtending angles $\theta$, $\theta'$ and $\beta$.
Equations~(\ref{phiphiprimedef}) are the law of cosines for spherical
triangles, applied to the interior angles $\phi$ and $\eta$, as shown
in the figure.  As for the third interior angle, we define $\kappa$ as
the opening angle of the lune (the shaded area), as illustrated in the
figure.  Then it is easy to show that the third interior angle of the
spherical triangle at vertex $\ahat$ is $\pi-\kappa$. For this angle
the law of cosines gives
	\begin{equation}
	\cos(\pi-\kappa) 
	=\frac{\cos\beta-\cos\theta\cos\theta'}
	{\sin\theta\sin\theta'}=-\cos\kappa,
	\label{kappadef}
	\end{equation}
determining $\kappa$ also uniquely in the interval $[0,\pi]$.

We define $\Phi_d$ as one half of the area of the shaded lune seen in
Fig.~\ref{intersect}, which is also the $F_4$-type generating function
of the transformation $(\theta,\phi)\to(\theta',\phi')$ (Sokolovski
and Connor\cite{SokolovskiConnor99}).  Then we have
	\begin{eqnarray}
	\Phi_d &=& J(\kappa - \phi\cos\theta - \eta\cos\theta')
	\nonumber \\
	   &=& J\kappa - J_z \phi -J_n \eta
	 =(j+\frac{1}{2})\kappa - m\phi -m'\eta,
	\label{Phiddef}
	\end{eqnarray}
where the final form applies if $J$, $J_z$ and $J_n$ take on their
quantized values.  The result is the sum of angular momentum quantum
numbers times dihedral angles, that is, the interior angles of the
spherical triangle formed by $(\zhat,\nhat,\ahat)$ are also the
interior dihedral angles of the tetrahedron or parallelepiped formed
by those vectors.  The analogy with the Ponzano-Regge formula is more
transparent if we use the exterior dihedral angles $\kappa$,
$\pi-\phi$ and $\pi-\eta$, so that
	\begin{equation}
	\Phi_d = (j+\frac{1}{2})\kappa + m(\pi-\phi) + m'(\pi-\eta)
	-(m+m')\pi,
	\label{altPhiddef}
	\end{equation}
in which the first three terms look like the sum (\ref{PhiPRdef}),
while the final term just produces a phase factor $(-1)^{m+m'}$ in the
asymptotic formula.

It is straightforward to prove (\ref{Phiddef}) by elementary geometry,
but another proof, based on symplectic reduction of Schwinger's
oscillator model of angular momentum\cite{BiedenharnvanDam65}
(essentially the Hopf fibration), leads to the following
generalization.  Let a polygon on the unit sphere be specified by
vertices $(\vhat_1, \ldots, \vhat_n)$ connected by arcs of small
circles, where the small circle proceeding from $\vhat_i$ to
$\vhat_{i+1}$ is obtained by rotating $\vhat_i$ about axis $\nhat_i$
by angle $\phi_i$, using the right-hand rule.  Also, let $\kappa_i$ be
the interior angle between the two small circles meeting at $\vhat_i$.
Then the solid angle of the interior of the polyhedron, defined as the
region to the left as we move along the small circles, is
	\begin{equation}
	\Omega = 2\pi - \sum_{i=1}^n [(\pi-\kappa_i)+(\vhat_i\cdot\nhat_i)
	\phi_i].
	\label{Omegaformula}
	\end{equation}
Special cases of this formula include the solid angle of a spherical
triangle (with sides that are great circles),
$\Omega=\kappa_1+\kappa_2+\kappa_3-\pi$, and (\ref{Phiddef}), for
which $n=2$, $\kappa_1=\kappa_2=\kappa$, and $\Phi_d = \Omega/2$.
Equation~(\ref{Omegaformula}) can also be derived as a special case of
the Gauss-Bonnet theorem.

The spherical triangle formed by $(\zhat,\nhat,\ahat)$ plays another
role.  We define $V_d$ as the volume of the parallelepiped spanned by
these three vectors, which can be written in a variety of ways,
	\begin{eqnarray}
	V_d &=& (\zhat\times\nhat)\cdot\ahat =
	\sin\beta\sin\theta\sin\phi \nonumber \\
	&=& \sin\beta\sin\theta'\sin\eta
	=\sin\theta\sin\theta'\sin\kappa,
	\label{Vddef}
	\end{eqnarray}
where we use the law of sines for the final three expressions.
One of these equalities is equivalent to the $y$-component of
(\ref{thetaphixfm}).  The square of $V_d$ is the determinant of the
Gram matrix formed by vectors $(\zhat,\nhat,\ahat)$,
	\begin{eqnarray}
	V_d^2&=&\det\left(\begin{array}{ccc}
	\zhat\cdot\zhat & \zhat\cdot\nhat & \zhat\cdot\ahat \\
	\nhat\cdot\zhat & \nhat\cdot\nhat & \nhat\cdot\ahat \\
	\ahat\cdot\zhat & \ahat\cdot\nhat & \ahat\cdot\ahat
	\end{array}\right)=
	\det\left(\begin{array}{ccc}
	1 & \cos\beta & \cos\theta \\
	\cos\beta & 1 &\cos\theta' \\
	\cos\theta & \cos\theta' &1
	\end{array}\right)
	\nonumber \\
	&=& 1 + 2\cos\beta\cos\theta\cos\theta' -\cos^2\beta
	-\cos^2\theta -\cos^2\theta'.
	\label{VdGram}
	\end{eqnarray}
Since we are working with the volume of the parallelepiped instead of
the volume of the tetrahedron, there is no factor of 6 in
(\ref{Vddef}) or of $6^2=36$ in (\ref{VdGram}).  The volume $V_d$
appears in the amplitude of the asymptotic formulas (\ref{djmmWKB})
and (\ref{djmmcfr}).

Including all the details (Maslov indices, phase conventions, etc),
the asymptotic expression for the $d$-matrix in the classically allowed
region is 
	\begin{equation}
	d^j_{mm'}(\beta) = A_d
	\cos\left(\Phi_d - \frac{\pi}{4}\right),
	\label{djmmWKB}
	\end{equation}
where the amplitude is
	\begin{equation}
	A_d = \frac{(-1)^{j-m'}} {\sqrt{(\pi/2)J|V_d|}}.
	\label{Adcar}
	\end{equation}
	
According to Ref.~\cite{Littlejohn90} the amplitude of the WKB
approximation for the $d$-matrix should be proportional to the inverse
square root of the Poisson bracket $\{J_z,J_n\}$, evaluated at the
intersection between the two orbits, $\Jvec=J\ahat$.  Indeed, using
(\ref{dPBdef}), we see that it is:
	\begin{eqnarray}
	\{J_z,J_n\} &=& (\zhat \times \nhat)\cdot \Jvec
	=JV_d
	= \sin\beta J_y \nonumber \\
	&=& J \sin\beta\sin\theta\sin\phi.
	\label{dAmpPB}
	\end{eqnarray}

The caustics of the $d$-matrices occur when the small circles
$J_z={\rm const}$ and $J_n={\rm const}$ are tangent, or, equivalently,
when the vectors $(\zhat,\nhat,\ahat)$ are linearly dependent so that
$V_d=0$.  Multiplying (\ref{VdGram}) by $J^2$, using
(\ref{thetathetaprimedef}) and setting the result to zero gives us the
equation of the caustic in $J_z$-$J_n$ space,
	\begin{equation}
	J_z^2 + J_n^2 -2J_zJ_n\cos\beta -J^2\sin^2\beta=0,
	\label{dcausticeqn}
	\end{equation}
an ellipse whose axes are oriented $45^\circ$ to the $J_z$-$J_n$ axes,
and whose semimajor and semiminor axes are $\sqrt{2}\cos(\beta/2)$,
$\sqrt{2}\sin(\beta/2)$.  An example is illustrated in
Fig.~\ref{dspots}; see also Fig.~1 of Ref.~\cite{Braunetal96}.  The
ellipse touches the boundary defined by the classical limits
(\ref{JzJnbounds}) at four points, creating four classically forbidden
regions labeled ABCD in Fig~\ref{dspots}.  The square of the volume
$V_d^2$ is negative in the classically forbidden regions, and $V_d$
itself is imaginary there.

Another point of view on the caustics is to hold $J_z$ and $J_n$
fixed, thereby fixing the sizes of the two small circles, and to vary
$\beta$, which moves the position of the small circle $J_n={\rm
const}$.  Then the small circles are tangent at the turning points
$\beta=\beta_1$ or $\beta_2$, where $0\le \beta_1 \le \beta_2 \le
\pi$, and where
	\begin{equation}
	\beta_1 = |\theta-\theta'|,
	\qquad
	\beta_2 = \min(\theta+\theta', 2\pi-\theta-\theta').
	\label{beta12def}
	\end{equation}
The classically allowed region is $\beta_1 \le \beta \le \beta_2$,
while the two classically forbidden regions are $0\le\beta\le\beta_1$
and $\beta_2\le\beta\le\pi$.

The four types of tangencies of the two small circles are illustrated
in Fig.~\ref{dcaustics}.  In all four parts of the figure, $T$ is the
caustic point (the point of tangency).  In part A we are at the upper
turning point $\beta=\beta_2$, because if $\beta$ decreases we obtain
two intersection points and are in the classically allowed region.  In
fact, this is the case $\beta_2=\theta+\theta'<\pi$.  Or if we hold
$\beta$ fixed but decrease either $J_z$ or $J_n$, again we enter the
classically allowed region, since one or the other of the two small
circles expands and the tangency develops into two intersection
points.  Thus part A of Fig.~\ref{dcaustics} corresponds to the corner
A of Fig.~\ref{dspots}.  In part B of Fig.~\ref{dcaustics} we are at
the lower turning point $\beta=\beta_1=\theta-\theta'>0$, since if
$\beta$ increases we move into the classically allowed region.  The
same happens if we hold $\beta$ fixed and either increase $J_z$ or
decrease $J_n$, so this corresponds to corner B of Fig.~\ref{dspots}.
In part C of Fig.~\ref{dcaustics} we are at the lower turning point
$\beta=\beta_1=\theta'-\theta>0$, which corresponds to corner C of
Fig.~\ref{dspots} since we enter the classically allowed region if
either $J_n$ increases or $J_z$ decreases.  Finally, in part D of
Fig.~\ref{dcaustics} we are at the upper turning point $\beta_2 =
2\pi-\theta-\theta'<\pi$, which corresponds to corner D of
Fig.~\ref{dspots} since we enter the classically allowed region if
either $J_z$ or $J_n$ increases.

The four types of tangencies of orbits for the $6j$-symbol,
illustrated in Fig.~\ref{caustics}, are topologically identical to the
four types for the $d$-matrices, illustrated in Fig.~\ref{dcaustics}.
Similarly, the four classically forbidden regions of the $6j$-symbol,
illustrated in Fig.~\ref{spots}, are in one-to-one correspondence with
the four classically forbidden regions of the $d$-matrices,
illustrated in Fig.~\ref{dspots}.  Comparing Figs.~\ref{spots} and
\ref{dspots}, we see that the labelings of the corners by ABCD are not
the same; but this is because the point on the $d$-sphere of maximum
$J_n$, namely, the point in the direction $\nhat$, corresponds to the
point on the $6j$-sphere of minimum $J_{23}$.  If the $J_{23}$ axis in
Fig.~\ref{spots} had been drawn increasing downward instead of upward,
the labels on all four corners (classically forbidden regions) of both
Fig.~\ref{spots} and Fig.~\ref{dspots} would coincide.

Referring to Fig.~\ref{dspots}, if we hold $\beta$ fixed and vary
$J_z$ or $J_n$, moving from the interior of the ellipse (the
classically allowed region) to the boundary (the caustic), then all the
angles $\phi$, $\eta$ and $\kappa$ approach either 0 or $\pi$,
depending on which segment ABCD of the boundary (the caustic curve) we
approach.  The values of these angles on the caustics are summarized
in Table~\ref{dtable}.  For uniformity of notation, we write
$\alpha_i$, $i=1,2,3$ for $\kappa$, $\phi$ and $\eta$, as indicated in
the table, and similarly we write $k_i$, $i=1,2,3$ for $j$, $-m$,
$-m'$, where the signs are the same as in the three terms of the expression
(\ref{Phiddef}) for $\Phi_d$.  Also shown in the table is the integer
$\nu_d$ for the four classically forbidden regions, defined by
	\begin{equation}
	\nu_{d}={\mathop{\sum_i}\nolimits}' k_i = {\rm integer},
	\label{nuddef}
	\end{equation}
where the sum is only taken over $i$ such that $\alpha_i=\pi$.  The
definition is similar to that of $\nu_{6j}$ in (\ref{nudef}), and used
for a similar purpose, that is, the asymptotic form of the
$d$-matrices in the classically forbidden regions carries a phase
$(-1)^{\nu_d}$, effectively due to the analytic continuation of
$\Phi_d$.  

\begin{table}
\begin{tabular}{|c|c|c||c|c|c|c|}\hline
 $i$ & $\alpha_i$ & $k_i$ & A & B & C & D \\
\hline\hline
1 & $\kappa$ & $j$ & 0 & $\pi$ & $\pi$ & 0\\
2 & $\phi$ & $-m$ & 0 & 0 & $\pi$ & $\pi$ \\
3 & $\eta$ & $-m'$ & 0 & $\pi$ & 0 & $\pi$ \\
\hline\hline
&& $\nu_d$ & 0 & $j-m'$ & $j-m$ & $-m-m'$ \\
\hline
\end{tabular}
\caption{\label{dtable} Values of the angles $\kappa$, $\phi$ and
$\eta$ on caustics of type ABCD, also integer $\nu_d$ for four caustic
types.}
\end{table}

The angles $\alpha_i$ are extended into the classically forbidden
region in a manner exactly like that used for the $\psi_i$ in the case
of the $6j$-symbol, as explained below (\ref{nudef}).  That is, if
$\alpha_i=0$ on the segment of the caustic curve adjacent to a given
classically forbidden region, then we define
$\alpha_i=i{\bar\alpha}_i$, where
${\bar\alpha_i}=\cosh^{-1}(\cos\alpha_i)$ is real and positive; while
if $\alpha_i=\pi$ on the caustic curve, then we define
$\alpha_i=\pi+i{\bar\alpha_i}$, so that
${\bar\alpha_i}=-\cosh^{-1}(-\cos\alpha_i)$ is real and negative.  In
the classically forbidden regions, the quantities $\cos\alpha_i$,
given by (\ref{phiphiprimedef}) and (\ref{kappadef}), lie outside the
interval $[-1,-1]$.  

We now define a quantity related to the analytic continuation of
$\Phi_d$ into the classically forbidden regions,
	\begin{equation}
	{\bar\Phi}_d = \sum_{i=1}^3 k_i {\bar\alpha}_i
	=\sum_{i=1}^3 k_i \sign(\cos\alpha_i)
	\cosh^{-1}|\cos\alpha_i|.
	\label{barPhiddef}
	\end{equation}
In spite of the absolute value and sign functions, ${\bar\Phi}_d$ is
smooth over any given classically forbidden region.  This is important
for root finders that rely on smoothness, such as the Newton-Raphson
method.  The quantity ${\bar\Phi}_d$ is zero on the caustic boundary,
and real and negative as we move into classically forbidden regions B
and C, and real and positive as we move into classically forbidden
regions A and D.  These are the same rules as for ${\bar\Phi}_{\rm
PR}$.  The $d$-matrices decrease exponentially as we move into any
classically forbidden region, a behavior that is captured by
$\exp(-|{\bar\Phi}_d|)$ in all cases.

Finally, the asymptotic expression for the $d$-matrices in the
classically forbidden region is
	\begin{equation}
	d^j_{mm'} = A_d
	\exp(-|{\bar\Phi}_d|),
	\label{djmmcfr}
	\end{equation}
where
	\begin{equation}
	A_d = \frac{(-1)^{j-m'+\nu_d}}{2\sqrt{(\pi/2) J |V_d|}}.
	\label{Adcfr}
	\end{equation}

\section{The Uniform Approximation}
\label{unifappx}

\subsection{Remarks on Uniform Semiclassical Ap\-prox\-i\-ma\-tions}
\label{remarksuniform}

The traditional method of constructing uniform semiclassical
approximations, the ``method of comparison equations,'' is reviewed by
Berry and Mount\cite{BerryMount72}, with citations to earlier
literature.  In this method one takes a one-dimensional Schr\"odinger
equation (a second-order differential equation in $x$) and performs a
coordinate transformation $X=X(x)$ to create a new Schr\"odinger
equation in $X$ which, after the neglect of terms of order $\hbar^2$,
becomes a standard, solvable equation.  The most common standard or
``comparison'' equations in practice are the differential equations for
Airy or Weber (parabolic cylinder) functions.

Since both the $6j$-symbols and the $d$-matrices satisfy second-order
difference equations, it is likely that a kind of discrete version of
the method of comparison equations, along the lines of the discrete
WKB theory used by Schulten and Gordon\cite{SchultenGordon75} and
refined by Braun\cite{Braun93}, could be used to construct a uniform
approximation for the $6j$-symbol.  We have not constructed our
approximation in this way, however, and if we had, it is likely that
we would have missed much of the geometry discussed above.  Also, that
approach would have produced a uniform approximation for the
$6j$-symbol only for a fixed value of $j_{12}$ or $j_{23}$, not over
the whole range of both variables as we have done here (it would not
have produced a uniform approximation in terms of $d$-matrices).

Hiding slightly beneath the surface of the usual method of comparison
equations is a transformation between the phase spaces of the original
problem and the standard problem.  This transformation is $X=X(x)$ and
$P=(dx/dX)p$, where the first part is the coordinate transformation
used in the method (a ``point transformation'') and the second part is
the usual lift of a point transformation into a canonical
transformation.  The second or momentum equation can also be written
$p\,dx=P\,dX$, which by integration gives
	\begin{equation}
	s(x)=S(X),
	\label{actionsequal}
	\end{equation}
where $s$ and $S$ are the actions of the original problem and the
standard problem, respectively.  In fact, this equation (the equality
of the actions) specifies the coordinate transformation $X=X(x)$.  The
geometry is illustrated in Fig.~\ref{morse}, in which the method of
comparison equations is used to map a quantized curve of a nonlinear
oscillator (the Morse oscillator, part (a) of the figure) into a
quantized curve of a standard problem (the harmonic oscillator, part
(b) of the figure).  Equation~(\ref{actionsequal}) implies the
equality of the shaded areas in the figure, which in turn determines
the function $X(x)$.  In the figure, $X_0 = X(x_0)$.

The function $X(x)$ can be analytically continued to the classically
forbidden regions where $p$, $P$, $s$ and $S$ all become complex, but
the transformation $X=X(x)$ is real and in fact forms a single, smooth
(usually analytic) coordinate transformation throughout both
classically allowed and forbidden regions.  See Fig.~2 of
Ref.~\cite{CargoLittlejohn02} for a plot of the function $X(x)$ in one
example, which shows its completely smooth behavior as one passes from
classically allowed to classically forbidden regions.  The equation
$s(x)=S(X)$ is obvious in a sense: actions are areas, and areas are
preserved by canonical transformations.

We have been interested in the generalization of the method of
comparison equations to a class of canonical transformations that is
larger than the point transformations.  Since canonical
transformations are the semiclassical representatives of unitary
transformations, the idea is to carry out a unitary transformation on
the original problem such that the transformed problem has a standard
form, to within errors of order $\hbar^2$, where the choice of unitary
transformation is guided by geometrical criteria in the classical
phase space.  The standard problem is sometimes referred to as a
``quantum normal form.''  Examples of quantum normal form calculations
and an interesting perspective on Bohr-Sommerfeld or torus
quantization may be found in Refs.~\cite{Cargoetal05a,Cargoetal05b}.
In those references only the problem of determining eigenvalues is
considered, but in the present application we are interested in the
transformation of the wave functions, a problem that involves extra
features.  As it turns out, it is not only necessary to carry out a
unitary transformation (which at the classical level maps a pair of
orbits into another pair that are in standard form), but also a
certain nonunitary transformation (to make the densities of
probability on the orbits come out in standard form).

In the usual method of comparison equations, the uniform approximation
for the exact solution $\psi(x)$ is given by
	\begin{equation}
	\psi(x) \approx \frac{a(x)}{A(X)} \Psi(X),
	\label{psiPsi}
	\end{equation}
where $\Psi(X)$ is the standard solution of the standard problem, and
$a(x)$ and $A(X)$ are the amplitudes of the two semiclassical
approximations, $a(x)$ for the original problem and $A(X)$ for the
standard problem.  Both amplitudes diverge at the caustics, but their
ratio has a definite limit and in fact is smooth everywhere across
both classically allowed and forbidden regions.

Similarly, it turns out that the uniform approximation for the
$6j$-symbol in terms of $d$-matrices is given by
	\begin{equation}
	\left\{ \begin{array}{ccc}
	j_1 & j_2 & j_{12} \\
	j_3 & j_4 & j_{23}
	\end{array}\right\} \approx (-1)^{\nu_{\rm ex}}
	\frac{A_{\rm PR}}{A_d} d^j_{mm'}(\beta),
	\label{theuniformappx}
	\end{equation}
where $A_{\rm PR}$ is given by (\ref{APRcar}) or (\ref{APRcfr}) and
$A_d$ by (\ref{Adcar}) or (\ref{Adcfr}) in the classically allowed or
forbidden regions, respectively, and where $(-1)^{\nu_{\rm ex}}$ is an extra
phase defined in (\ref{nuexdef}).  In other words, the uniform
approximation for the $6j$-symbol is of the same form
(\ref{psiPsi}) that emerges from the method of comparison
equations, apart from an extra phase. 

\subsection{The Details of the Uniform Approximation}
\label{detailsuniform}

At the heart of the new uniform approximation is a smooth,
area-preserving map between the $6j$-sphere and the $d$-sphere that
is parameterized by fixed, quantized values of $J_{12}$ and $J_{23}$,
related to quantum numbers $j_{12}$ and $j_{23}$ by (\ref{Jidef}).
The area of the $6j$-sphere and that of the $d$-sphere must be equal,
which implies
	\begin{equation}
	D=2j+1,
	\label{jddef}
	\end{equation} 
where $D$ is given by (\ref{dimZdef}) or (\ref{Ddef}).  Thus the value
of the parameter $j$ of the $d$-matrix in (\ref{theuniformappx}) is
determined.  

The quantum numbers $j_{12}$ and $j_{23}$ determine a specific pair of
orbits on the $6j$-sphere, the small circle
$J_{12}=j_{12}+\frac{1}{2}$, and the oval $J_{23}=j_{23}+\frac{1}{2}$.
The map is required to map the small circle
$J_{12}=j_{12}+\frac{1}{2}$, a quantized orbit, onto a small circle
$J_z={\rm const}$ on the $d$-sphere.  Because area is preserved, the
small circle on the $d$-sphere must also be quantized, and contain the
same area about the north pole as the original small circle on the
$6j$-sphere.  That is, we must have $j_{12,{\rm max}} - j_{12} = j-m$,
or,
	\begin{equation}
	m=j_{12} - j_{12,{\rm avg}},
	\label{mj12eqn}
	\end{equation}
where $j_{12,{\rm avg}}=(j_{12,{\rm min}} + j_{12,{\rm max}})/2$.
This determines the quantum number $m$ in (\ref{theuniformappx}).  We
only require our map to map this specific small circle of constant $J_{12}$
on the $6j$-sphere onto the corresponding small circle on the
$d$-sphere; other small circles of constant $J_{12}$ on the
$6j$-sphere, for other values of $J_{12}$, are {\it not} mapped to
small circles of constant $J_z$ on the $d$-sphere.  

Similarly, we require the map to map the quantized oval $J_{23} =
j_{23}+\frac{1}{2}$ on the $6j$-sphere onto a small circle on the
$d$-sphere that is centered about some direction $\nhat$ that lies on
the semicircle $J_y=0$, $J_x>0$ on the $d$-sphere.  The direction
$\nhat$ is a function of the angle $\beta$, which will be specified
momentarily.  Because area is preserved, the new small circle on the
$d$-sphere will be a quantized orbit $J_n=m'$, enclosing the same area
about the axis $\nhat$ as the oval $J_{23}=j_{23}+\frac{1}{2}$ encloses
about the point $J_{23} = J_{23,{\rm min}}$.  Since the minimum of
$J_{23}$ corresponds to the maximum of $J_n$, the quantum number $m'$
satisfies $j_{23}-j_{23,{\rm min}} = j-m'$, or,
	\begin{equation}
	m'=j_{23,{\rm avg}} - j_{23}.
	\label{mprimej23eqn}
	\end{equation}
This determines the quantum number $m'$ in (\ref{theuniformappx}).  We
only require our map to map this specific oval of constant $J_{23}$ on
the $6j$-sphere onto the corresponding small circle on the $d$-sphere;
other ovals of constant $J_{23}$ on the $6j$-sphere, for other values
of $J_{23}$, are {\it not} mapped to small circles of constant $J_z$
on the $d$-sphere. 

Finally, the parameter $\beta$ is determined by requiring that the
area of the lune on the $6j$-sphere should equal the area of the lune
on the $d$-sphere.  Effectively, we rotate the small circle $J_n=m'$
until the two areas are equal.  This is in the classically allowed
region; in the classically forbidden region, the analytic
continuations of the areas on the two spheres are set equal.  In this
way, if the $6j$-symbol is in the classically allowed region, then so
is the $d$-matrix, and vice versa.  This condition is the analog of
(\ref{actionsequal}) in the standard method of comparison equations.

We take the classically allowed region first.  We shall show elsewhere
that the Ponzano-Regge phase is related to the area of the lune on the
$6j$-sphere by
	\begin{equation}
	\Phi_{\rm PR} = \frac{1}{2}(\hbox{\rm Area of lune}) +
	\Phi_0,
	\label{PhiPRarea}
	\end{equation}
where $\Phi_0$ is an extra phase  that
is related to the topology of loops in the phase space of
Schwinger's oscillators, as discussed in Ref.~\cite{Aquilantietal07}.
Without going into this, we can determine $\Phi_0$ by evaluating
both the area of the lune and $\Phi_{\rm PR}$ at any point in the
classically allowed region or on the caustic curve, as in
Fig.~\ref{spots}, since $\Phi_{\rm PR}$ is a continuous function of
position inside and on that curve.  A point on segment A of the
caustic boundary is convenient, since the area of the lune vanishes
there, as shown by part A of Fig.~\ref{caustics}.  In this way we find
	\begin{equation}
	\Phi_0 = (J_1+J_2+J_3+J_4+J_{12} -J_{12,{\rm max}})\pi
	= (\nu_{\rm ex} +\frac{3}{2})\pi,
	\label{extraphase}
	\end{equation}
where
	\begin{equation}
	\nu_{\rm ex} = j_1+j_2+j_3+j_4+j_{12}-j_{12,\rm{max}}.
	\label{nuexdef}
	\end{equation}
Note that $\nu_{\rm ex}$ is an integer.  As a check one can evaluate
$\Phi_{\rm PR}$ and the area of the lune at other points on the
caustic boundary (segments BCD), and see that the answer for $\Phi_0$
agrees with (\ref{extraphase}).

As for the area of the lune on the $d$-sphere, it is a function of
$\beta$ and is twice $\Phi_d$, given by (\ref{Phiddef}).  Altogether, the
equation that must be solved for $\beta$ in the classically allowed
region is
	\begin{equation}
	\Phi_{\rm PR} = \Phi_d(\beta)+\Phi_0.
	\label{betaeqncar}
	\end{equation}
In spite of the complications arising from the extra term $\Phi_0$,
the geometrical meaning of (\ref{betaeqncar}) is simple:  the areas of
the lunes on the two spheres are equal. 

Taking cosines, (\ref{betaeqncar}), (\ref{extraphase}) and
(\ref{nuexdef}) imply
	\begin{equation}
	\cos(\Phi_{\rm PR}+\frac{\pi}{4}) =
	\cos(\Phi_d +\nu_{\rm ex}\pi + \frac{3\pi}{2} + \frac{\pi}{4})
	=(-1)^{\nu_{\rm ex}} \cos(\Phi_d - \frac{\pi}{4}),
	\label{cosPhieqn}
	\end{equation}
which explains the extra phase in (\ref{theuniformappx}) as well as
the opposite signs on the $\pi/4$ in the asymptotic formulas
(\ref{PRappx}) and (\ref{djmmWKB}).	

In the classically forbidden region, the analytic continuations of the
areas of the two lunes become complex, but their real parts are
constant in any given region (ABCD), so only the imaginary parts need
be equated.  In this case, the condition is
	\begin{equation}
	{\bar\Phi}_{\rm PR} = {\bar\Phi}_d(\beta),
	\label{betaeqncfr}
	\end{equation}
where ${\bar\Phi}_{\rm PR}$ is given by (\ref{barPhiPRdef}) and
${\bar\Phi}_d$ by (\ref{barPhiddef}).  Since both ${\bar\Phi}_{\rm
PR}$ and ${\bar\Phi}_d$ vanish on the caustic, the condition
(\ref{betaeqncfr}) implies that if the $6j$-symbol is on a caustic,
then so is the $d$-matrix (as it should be).

To find the root of (\ref{betaeqncar}) in the classically allowed
region it helps to know the values of the areas of the lunes on the
$d$-sphere at the two turning points $\beta_1$ and $\beta_2$, defined
by (\ref{beta12def}).  Call the half areas of the $d$-lunes at
these two turning points ${\cal A}_1$ and ${\cal A}_2$.  Then we have
	\begin{equation}
	{\cal A}_1 = [j+\frac{1}{2}-\max(m,m')]\pi,
	\qquad
	{\cal A}_2 = \max(0,-m-m')\pi.
	\label{areatps}
	\end{equation}
From these we compute an initial estimate for the root based on linear
interpolation,
	\begin{equation}
	\beta_0=\beta_1 + \frac{{\cal A}_1 -\Phi_{\rm PR}+\Phi_0}
	{{\cal A}_1-{\cal A}_2}(\beta_2-\beta_1),
	\label{beta0car}
	\end{equation}
after which a Newton-Raphson iteration converges to the actual root in
all the cases we have examined, without taking iterations outside the
interval $[\beta_1,\beta_2]$.  We have no proof that this is always
true, however.

As for the classically forbidden regions, there are two of them, $0 <
\beta < \beta_1$ and $\beta_2 < \beta < \pi$, in which we must solve
(\ref{betaeqncfr}).  Here we cannot use linear interpolation to find an
initial estimate of the root since $\Phi_d$ is not defined at
$\beta=0$ or $\beta=\pi$.  Instead we have simply taken 
	\begin{equation}
	\beta_0 = \left\{\begin{array}{r@{\quad}l}
	\frac{1}{2}\beta_1, & 0<\beta<\beta_1, \\
	\frac{1}{2}(\beta_2+\pi), & \beta_2<\beta<\pi,
	\end{array}\right.
	\label{beta0cfr}
	\end{equation}
and then applied a Newton-Raphson iteration.  Occasionally this takes
us outside the given classically forbidden region, whereupon we have
reset the value of $\beta$ to a point inside the region, using a
simple prescription.  After this, the Newton-Raphson converges to the
root in all the cases we have examined.  Our algorithm has proven
satisfactory for our exploratory studies, but in more serious work the
root finder will require more careful attention.

The Newton-Raphson method requires us to know the derivatives
$d\Phi_d/d\beta$ or $d{\bar\Phi}_d/d\beta$.  These also
appear in the role that $\Phi_d$ plays as a generating function, and
are equal to the momentum $p_\beta$ of the rigid body, of which
$d^j_{mm'}(\beta)$ is an eigenfunction (on the group manifold $SU(2)$).
The derivatives are given by
	\begin{equation}
	\frac{d\Phi_d}{d\beta} \quad {\rm or} \quad
	\frac{d{\bar\Phi}_d}{d\beta} = 
	-\frac{J|V_d|}{\sin\beta}.
	\label{pbetadef}
	\end{equation}
This applies at the intersection $\ahat$ (not $\ahat'$) in
Fig.~\ref{intersect}, or at its analytic continuation as specified by
the definitions of the angles ${\bar\alpha}_i$ at the end of
Sec.~\ref{SCMdmatrices}.  It is convenient in using this formula to
avoid calculating the sines of $\kappa$, $\phi$ or $\eta$, which
become complex in the classically forbidden region; this can be done
if $|V_d|$ is evaluated by taking the square root of the expression in
(\ref{VdGram}).

The value of $\beta$ is smooth and well behaved as we cross from the
classically allowed to forbidden regions, just as is the function
$X(x)$ in the method of comparison equations.  This is illustrated in
Fig.~\ref{beta}, which uses the same parameters as Fig.~\ref{spots}.
The heavy line in the figure is the same caustic line as in
Fig.~\ref{spots}, and the light lines are contours of $\beta$, labeled
in degrees.

Finally, we remark that the ratio of the amplitudes $A_{\rm PR}/A_d$
in (\ref{theuniformappx}) has the form $\infty/\infty$ as the caustic
is approached, so a well designed numerical implementation of that
formula would give special treatment to a small region around the
caustic, where l'Hospital's rule would be used to avoid numerical
difficulties. 

\subsection{Numerical Results}
\label{numericalresults}

Figure~\ref{errors} shows the results of numerical tests of the new
uniform approximation, with comparison with the Ponzano-Regge
approximation.  In the figure errors are plotted as a function of
$j_{12}$ for the $6j$-symbols
	\begin{equation}
	\left\{ \begin{array}{ccc}
	39/2 & 23 & j_{12} \\
	17/2 & 20 & 47/2
	\end{array}\right\} 
	\quad {\rm and} \quad
	\left\{ \begin{array}{ccc}
	156 & 184 & j_{12} \\
	68 & 160 & 188
	\end{array}\right\},
	\label{error6js}
	\end{equation}
in part~(a) and part~(b) of the figure, respectively.  The values of
the five fixed $j$'s in part~(b) are 8 times those in part~(a).  The
plots show the absolute value of the difference between the exact
$6j$-symbol and the approximate value.  In both parts of the figure,
the curve labeled PR is the error of the Ponzano-Regge approximation,
while that labeled U is the error of the uniform approximation.  The
error of the Ponzano-Regge approximation is large near the caustics,
as expected, while the error of the uniform approximation is fairly
flat throughout the classically allowed region and up to the caustics.
The error of both approximations falls rapidly to zero in the
classically forbidden regions, as of course does the exact
$6j$-symbol.

We computed the exact $6j$-symbol with extended precision integer
arithmetic.  Recently other calculations have used extended precision
floating point arithmetic to study the Wigner $3nj$-symbols and their
asymptotic properties\cite{Andersonetal08,Ragnietal09}.  Extended
precision is required when summing series with alternating (or
variable) sign; often the desired sum is an exponentially small
residue left when much larger terms nearly cancel.  With floating
point arithmetic one must carry enough extra precision so that enough
remains after the subtractions; with integer arithmetic the answer is
exact, but the cancellations are still present (hence higher precision
is carried on intermediate results).

In the classically allowed region the error is oscillatory, and it is
possible for the Ponzano-Regge error to be less than that of the
uniform approximation simply because it accidentally happens to fall
near a zero of the cosine function.  One such descending spike near
$j_{12}=140$ can be seen in part~(b) of the figure.  It is clear,
however, that a fair comparison of the errors in the classically
allowed region must use the amplitude of the oscillatory function and
ignore the oscillations.  By this measure the error of the uniform
approximation in Fig.~\ref{errors} is approximately 30 times smaller
than that of the Ponzano-Regge approximation in the center of the
classically allowed region, and gets better as we approach the
caustics.  This ratio is nearly the same in parts~(a) and (b),
indicating that both errors scale in the same way with $j$.  In the
classically forbidden region the errors can be compared directly,
without removing any oscillatory factor, and again for the values used
the figure shows that the error in the uniform approximation is
smaller than that in the Ponzano-Regge approximation.

The error term for the Ponzano-Regge approximation is unknown, as is
that for the uniform approximation, so there is no theory by which the
errors can be compared.  We would expect, however, on general grounds
that the two error terms should scale the same with $j$, a conclusion
that is supported by the numerical evidence.  That the ratio between
the errors should be as small as seen in (\ref{error6js}) and
Fig.~\ref{errors} was a surprise to us, and we have no explanation for
it.  The values of the $j$'s chosen in that example were essentially
random, but when we try other ``randomly chosen'' values of the $j$'s
we get similar plots.  If we systematically search for $j$ values such
that the error of the uniform approximation is as unfavorable as
possible relative to that of the Ponzano-Regge approximation in the
classically allowed region, we find cases such as
	\begin{equation}
	\left\{ \begin{array}{ccc}
	44 & 40 & j_{12} \\
	20 & 24 & 28
	\end{array}\right\},
	\label{sameerrs6j}
	\end{equation}
which gives the error plots in Fig.~\ref{sameerrs}.  In this
case the two errors are comparable for an extended range of $j_{12}$.
We have found no cases in which the uniform approximation is much
worse than the Ponzano-Regge approximation in the classically allowed
region.  

There is the question of when the uniform approximation is worst in an
absolute sense.  We define the relative error as the difference
between the approximation and the exact value, divided by a reference
value.  In the classically forbidden region, the reference value is
the absolute value of the exact value.  Inside the maximum part of the
Airy function lobe around the turning points the reference value is
the value of the Airy function, without any cosine modulation.
Elsewhere in the classically allowed region we take the reference
value to be the amplitude of the Ponzano-Regge approximation.

Using this definition of relative error, we have systematically
searched for $j$ values that make the relative error in the uniform
approximation largest.  We find that they occur in cases for which
$j_{12}=j_{23}=0$.  This can only happen when the other four $j$'s are
equal, so we have a $6j$-symbol of the form
	\begin{equation}
	\left\{ \begin{array}{ccc}
	j & j & 0 \\
	j & j & 0
	\end{array}\right\}.
	\label{worstcase}
	\end{equation}
For such symbols, the relative errors in both the uniform and the
Ponzano-Regge approximation actually increase with $j$, reaching
approximately 0.5 (uniform) or 1.1 (Ponzano-Regge) when $j=10$.  The
tetrahedra corresponding to $6j$-symbols of the form (\ref{worstcase})
have two small edges ($j_{12}$ and $j_{23}$, with lengths
$J_{12}=J_{23}=\frac{1}{2}$) with no vertex in common, as illustrated
in part~(a) of Fig.~\ref{badtetrahedra}.

The uniform approximation is asymptotic, so it is no surprise that it
does not work well for small quantum numbers such as those appearing
in (\ref{worstcase}).  Nevertheless it is interesting to see in more
detail why the approximation is not good.  Figure~\ref{badcase} shows
the $6j$-sphere for a symbol of the form (\ref{worstcase}).  The orbit
$j_{12}=0$ ($J_{12}=\frac{1}{2}$) is the small circle about the south
pole, while the orbit $j_{23}=0$ ($J_{23}=\frac{1}{2}$) is the curve
that ends at the south pole in a cusp.  The orbit $j_{23}=0$ is not
smooth at the south pole, and it cannot be deformed into a small
circle on the $d$-sphere by any smooth map.  Since the map is not
smooth, it has infinite derivatives, and the semiclassical
approximation breaks down.  In fact, in this case, even the
$6j$-sphere itself is not smooth.  That is, the $6j$-sphere is
obtained by symplectic reduction from a higher-dimensional phase
space, and it is always topologically a sphere.  But when the first
four $j$'s are all equal as in (\ref{worstcase}), the sphere is not
differentiable at the south pole.  In this case, it would be more
appropriate to think of a tear drop with its cusp at the south pole,
rather than a sphere.

Unlike the Ponzano-Regge approximation, the uniform approximation is
not invariant under all the symmetries of the $6j$-symbol.  It is
invariant under the three operations in which the upper and lower
elements of two columns are swapped, but not under the six
permutations of the columns.  (We have not tested the ``extra''
symmetries due to Regge\cite{BiedenharnLouck81}.)  Therefore when
$j\ne0$, we can permute columns in an unfavorable case such as
(\ref{worstcase}) to obtain a better approximation.  We have tested an
algorithm in which, before applying the uniform approximation, the
columns of the $6j$-symbol are permuted to place the column with the
largest minimum value in the third column.  Then when we search for
the worst case of the uniform approximation, we find that they occur
with symbols of the form
	\begin{equation}
	\left\{ \begin{array}{ccc}
	0 & 0 & 0 \\
	j & j & j
	\end{array}\right\},
	\label{3zeros6j}
	\end{equation}
in which three of the $j$'s are zero.  Symbols of this form correspond
to tetrahedra that look like part~(b) of Fig.~\ref{badtetrahedra}.  In
this case the relative error in the uniform approximation no longer
grows with $j$, rather it seems to approach a limit of about .075.

Note that if any $j$ in a $6j$-symbol vanishes, the symbol can be
evaluated trivially in closed form.  Thus, no approximation is needed
in the worst cases (\ref{worstcase}) and (\ref{3zeros6j}) that we have
examined.

\section{Conclusions}
\label{conclusions}

We have written computer codes that implement the uniform
approximations given in this paper, which document the algorithms and
check all their details.  We will make these programs available to any
interested parties.  

We will publish the derivation of the uniform approximation presented
in this paper in a future article, in which we will make explicit the
symplectic map between the $6j$-sphere and the $d$-sphere that
underlies it, as well as outline how the theory of quantum normal
forms leads to a uniform approximation in cases like this.

\section*{Acknowledgments}

It is a pleasure to offer this manuscript as a present to Enzo
Aquilanti on the occasion of his seventieth birthday, and to thank him
for suggesting the problem of uniform approximations to the
$6j$-symbol.  We would also like to thank him for years of friendship
and intellectual stimulation, which we hope and trust is only a
beginning.  

\appendix
\section{Constructing the tetrahedron}
\label{constucttetrahedron}

The $6j$-symbol specifies the lengths $J_i$ of the classical angular
momentum vectors $\Jvec_i$ but not their directions, so there is the
question of how the actual vectors can be constructed in three
dimensional space, satisfying the identities (\ref{Jsumvanishes}) and
(\ref{J12J23defs}).

Initially we assume that real vectors $\Jvec_i$ exist, and we define
	\begin{eqnarray}
	\Avec_1 &=& \Jvec_1, \nonumber \\
	\Avec_2 &=& \Jvec_1 + \Jvec_2 = \Jvec_{12}, \nonumber \\
	\Avec_3 &=& \Jvec_1 + \Jvec_2 + \Jvec_3 = -\Jvec_4,
	\label{Adefs}
	\end{eqnarray}
which are the three vectors running along the edges emanating from the
upper vertex in Fig.~\ref{tetrahedron}.  We arrange these vectors as
columns of a $3\times 3$ matrix $F$, and we let $G=F^TF$, where $T$
means transpose.  Then $G$ is the symmetric, nonnegative definite Gram
matrix of dot products,
	\begin{equation}
	G_{ij} = \Avec_i \cdot \Avec_j.
	\label{Gijdef}
	\end{equation}
By using the geometry of the three triangles spanned by the $\Avec_i$,
the components of $G$ can be found in terms of the lengths $J_i$,
	\begin{eqnarray}
	&G_{11} =  A_1^2 = J_1^2, \qquad
	G_{22} = A_2^2 = J_{12}^2, \qquad
	G_{33} = A_3^2 = J_4^2, \nonumber \\
	&G_{12} = G_{21} = \Avec_1 \cdot \Avec_2 =
		\frac{1}{2}(J_{12}^2 + J_1^2 -J_2^2), 
	\nonumber \\
	&G_{13} = G_{31} = \Avec_1 \cdot \Avec_3 =
	 	\frac{1}{2}(J_1^2 +J_4^2 -J_{23}^2), 
	\nonumber \\
	&G_{23} = G_{32} = \Avec_2 \cdot \Avec_3 =
		\frac{1}{2}(J_{12}^2 +J_4^2 -J_3^2).
	\label{Gijcalc}
	\end{eqnarray}
Alternatively, without making any assumptions about the existence of
the $\Jvec_i$, we can define $G$ in terms of the lengths $J_i$ by
(\ref{Gijcalc}).  Then there is the question of whether vectors
$\Avec_i$ exist such that (\ref{Gijdef}) is satisfied.

The diagonalization of $G$ is closely related to the singular value
decomposition of $F$.  The latter is $F=U DV^T$, where $U$ and $V$ are
real orthogonal matrices and $D$ is a real diagonal matrix, containing
the real singular values $d_i$ on the diagonal.  But this implies $G=V
D^2 V^T$, so $V$ is the orthogonal matrix that diagonalizes $G$ and
the eigenvalues of $G$ are $d_i^2$.  Therefore to find $F$ we first
construct $G$ by (\ref{Gijcalc}) and diagonalize it, obtaining $V$ and the
eigenvalues of $G$.  If these eigenvalues are all nonnegative, then their
square roots are the singular values, and the matrix $D$ is
determined.  This does not determine $U$, but that matrix amounts to
an overall rotation of the tetrahedron which is arbitrary anyway.  So
we can set $U$ to anything convenient, such as the identity.  Then
we have $F=DV^T$, and the vectors $\Avec_i$ can be obtained as the
columns of $F$.  From these we can find the $\Jvec$'s by inverting
(\ref{Adefs}) and using (\ref{J12J23defs}).  

Thus a real tetrahedron can be constructed if and only if the
eigenvalues of $G$ are nonnegative.  The tetrahedron is only
determined modulo overall rotations, proper and improper.  If we wish
the tetrahedron to have a definite handedness, we can perform a
spatial inversion, if necessary, to make the volume in (\ref{Vdef})
positive.  The spatial inversion is properly brought about by time
reversal, not parity, which does not change the sign of angular
momenta.  With this understanding, the final tetrahedron is determined
modulo proper rotations.

If any of the eigenvalues of $G$ are negative, then a real tetrahedron
does not exist.  It turns out that at most one eigenvalue of $G$
defined by (\ref{Gijcalc}) can be negative, assuming the triangle
inequalities on the $J_i$.  A proof of this fact is easily given by
considering the secular polynomial of $G$, whose coefficients can be
expressed in terms of the lengths of the $\Jvec$'s and the areas of
the triangles formed by them.  If $G$ has one negative eigenvalue,
then we can order the singular values so that the imaginary one is the
third one.  Then from $F=DV^T$ we see that the tetrahedron can be
constructed with complex vectors, in which the $x$- and $y$-components
of the $\Jvec$'s are real, and the $z$-components are purely
imaginary.

\pagebreak

\begin{figure}
\caption{\label{noweber} The $6j$-symbol as a function of $j_{12}$ for
$j_1=16$, $j_2=80$, $j_3=208$, $j_4=272$ and $j_{23}=276$.  Sticks are
the values of the $6j$-symbol, and the curve is the Ponzano-Regge
approximation.}
\end{figure}

\begin{figure}
\caption{\label{tetrahedron} A tetrahedron of positive volume with
conventional labeling of edges by angular momentum vectors.}
\end{figure}

\begin{figure}
\caption{\label{spots} The $J_{12}$--$J_{23}$ plane for
$j_1=\frac{9}{2}$, $j_2=3$, $j_3=\frac{11}{2}$, $j_4=6$.  The
classical bounds are $J_{12,{\rm min}}=\frac{3}{2}$, $J_{12,{\rm
max}}=\frac{17}{2}$, $J_{23,{\rm min}}=\frac{5}{2}$, $J_{23,{\rm
max}}=\frac{19}{2}$.  The dimension of the matrix $\langle j_{12}
\vert j_{23}\rangle$ is $D=\dim Z = 7$.  The point $J_{12}=5$,
$J_{23}=9$ ($j_{12}=\frac{9}{2}$, $j_{23}=\frac{17}{2}$) is very close
to the caustic line, but lies just inside.  The Ponzano-Regge
approximation is too large by a factor of 7 at this point.}
\end{figure}

\begin{figure}
\caption{\label{flat} A sequence of four flat tetrahedra, moving
around the caustic line of Fig.~\ref{spots} in a clockwise direction
from point Y.  The parameters are the same as in Fig.~\ref{spots}.
The numbers 1, 2 etc refer to vectors $\Jvec_1$, $\Jvec_2$, etc.}
\end{figure}

\begin{figure}
\caption{\label{dihedral} Definition of the interior dihedral angle
$\phi_{12}$.  The other interior dihedral angle $\phi_{23}$
is defined similarly.}
\end{figure}

\begin{figure}
\caption{\label{6jsphere} The phase space of the $6j$-symbol is a
sphere of radius $D/2$ in a space in which $(K_x,K_y,K_z)$ are
Cartesian coordinates.  To within an additive constant, $K_z$ is
$J_{12}$ and the azimuthal angle $\phi_{12}$ is the dihedral angle
of the tetrahedron.  Several curves of constant $J_{12}$ (small
circles) are shown.}
\end{figure}

\begin{figure}
\caption{\label{J23orbits} Curves of constant $J_{23}$ on the
$6j$-sphere.  The first view shows the north pole and the point
$J_{23}=J_{23,{\rm min}}$, and the second shows the south pole and the
point $J_{23}=J_{23,{\rm max}}$.}
\end{figure}

\begin{figure}
\caption{\label{carcfr} In part (a), the classically allowed region,
an orbit of constant $J_{12}$ intersects an orbits of constant
$J_{23}$.  The shaded area is the Ponzano-Regge phase, to within an
additive constant.  In part (b), the classically forbidden region, the
orbits do not intersect.}
\end{figure}

\begin{figure}
\caption{\label{caustics} Caustics occur when the curve $J_{12}={\rm
const}$ is tangent to the curve $J_{23}={\rm const}$.  The four types
of such tangencies are illustrated.}
\end{figure}

\begin{figure}
\caption{\label{euler} Definition of Euler angle $\beta$ and unit
vector $\nhat$.}
\end{figure}

\begin{figure}
\caption{\label{znorbits} Curves of constant $J_z$ and $J_n$ may
intersect in the classically allowed region (a), or not intersect in
the classically forbidden region (b).}
\end{figure}

\begin{figure}
\caption{\label{dspots} The square identifies the bounds on the
classical observables $J_z$ and $J_n$, while the spots indicate the
quantized values $J_z=m$, $J_n=m'$.  The ellipse is the caustic
curve.}
\end{figure}

\begin{figure}
\caption{\label{intersect} Vector $\ahat$ points to the intersection
of the $J_z$-orbit with the $J_n$-orbit, with $J_y>0$.  Vectors
$\zhat$, $\nhat$ and $\ahat$ define a spherical triangle, with
interior angles $\phi$, $\eta$ and $\kappa$.}
\end{figure}

\begin{figure}
\caption{\label{dcaustics} Caustics of the $d$-matrices occur when the
two small circles $J_z={\rm const}$ and $J_n={\rm const}$ are
tangent.  There are four possible configurations.}
\end{figure}

\begin{figure}
\caption{\label{morse} In the method of comparison equations, the
phase space of a nonlinear oscillator (a) is mapped into the phase
space of the harmonic oscillator (b).  The function $X(x)$ is
determined by the equality of areas; for example, in the figure the
shaded areas are equal, and $X_0 = X(x_0)$.}
\end{figure}

\begin{figure}
\caption{\label{beta} Contours of $\beta$, the root of
(\ref{betaeqncar}) or (\ref{betaeqncfr}), in the $J_{12}$-$J_{23}$
plane.  Parameters are same as in Fig.~\ref{spots}, but quantized
values are omitted.}
\end{figure}

\begin{figure}
\caption{\label{errors} Absolute value of the error of the
Ponzano-Regge approximation (PR) and of the uniform
approximation (U) as a function of $j_{12}$ for values of
the other five $j$'s shown in (\ref{error6js}).  The error is defined
as the difference between the approximate value and the exact value.}
\end{figure}

\begin{figure}
\caption{\label{sameerrs} Comparison of Ponzano-Regge (PR) and uniform
(U) errors as a function of $j_{12}$ for the $6j$-symbol
(\ref{sameerrs6j}).  Dotted curve is the error in the uniform
approximation.}
\end{figure}

\clearpage  

\begin{figure}
\caption{\label{badtetrahedra} The uniform approximation is worst for
tetrahedra that look like part~(a), where the small edges are $j_{12}$
and $j_{23}$.  If columns are permuted to optimize the approximation,
the worst case involves tetrahedra that look like part~(b), where
three edges are small.}
\end{figure}

\begin{figure}
\caption{\label{badcase} The orbits $j_{12}=0$ is a small circle about
the south pole, while the orbit $j_{23}=0$ has a cusp at the south
pole.  The shaded area is the ``lune.''}
\end{figure}

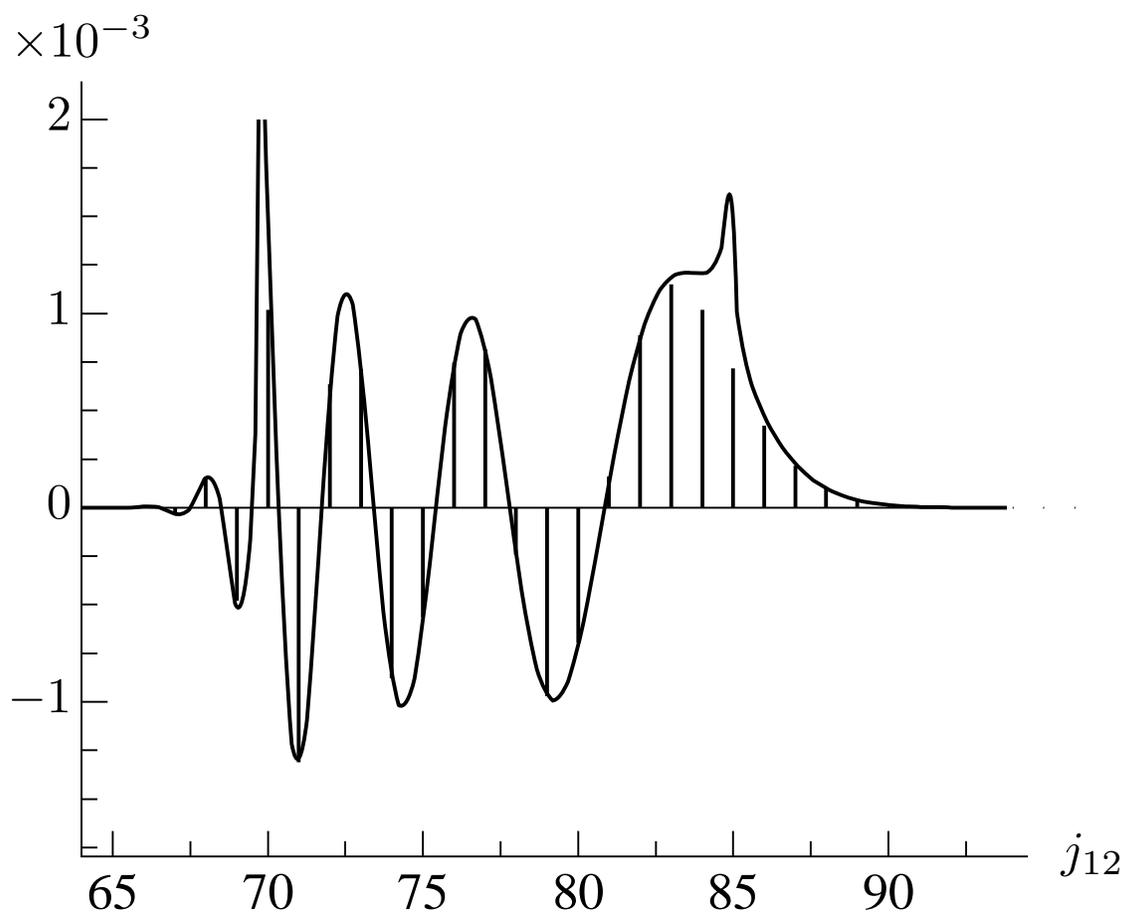
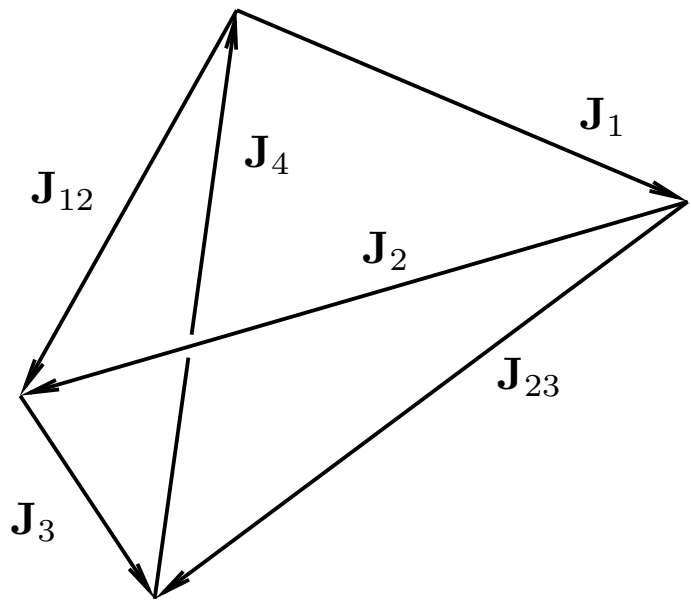
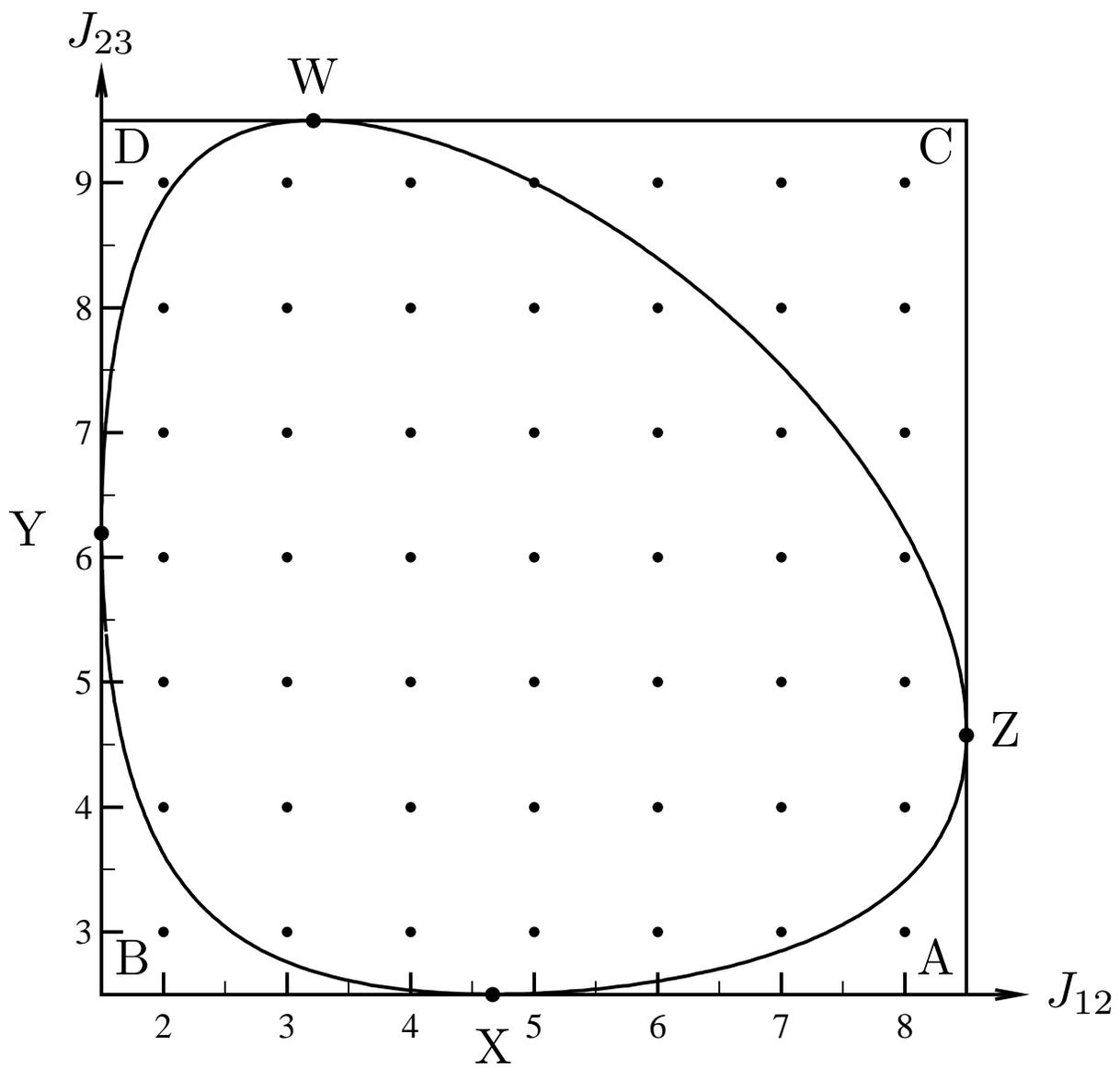
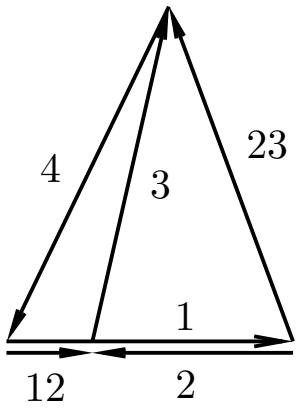
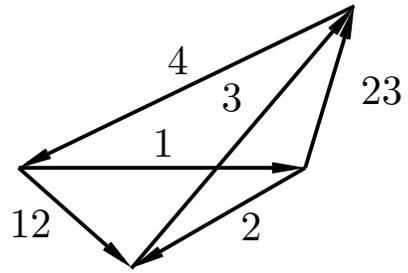
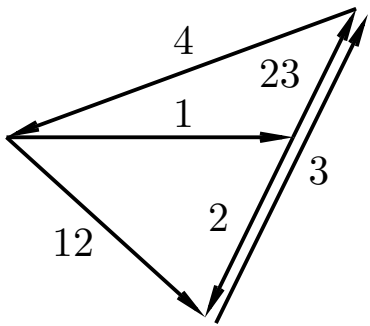
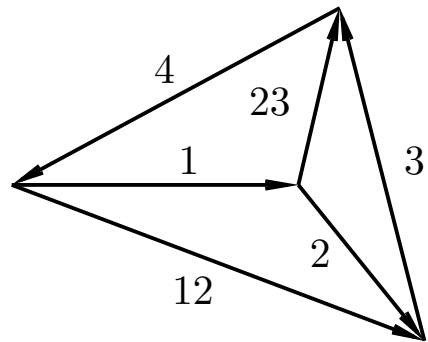
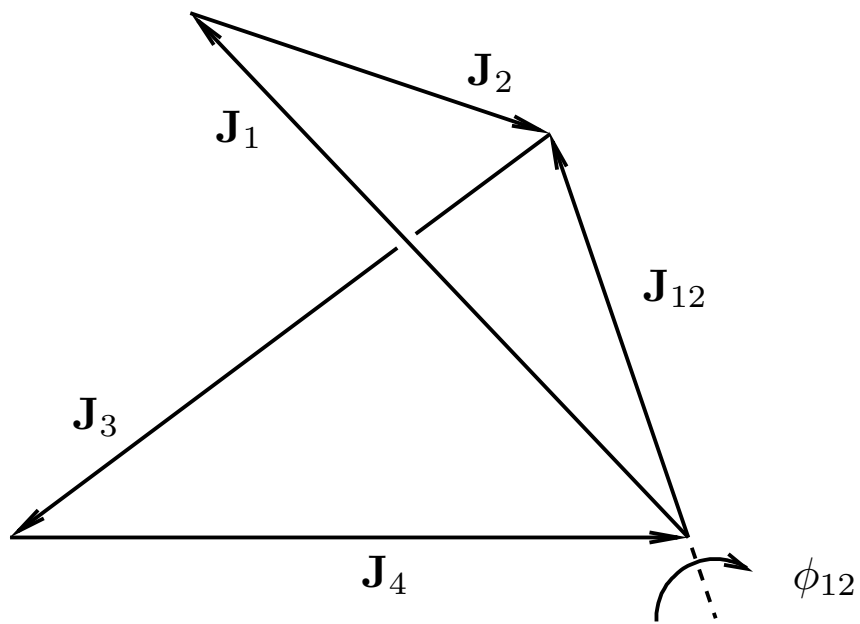
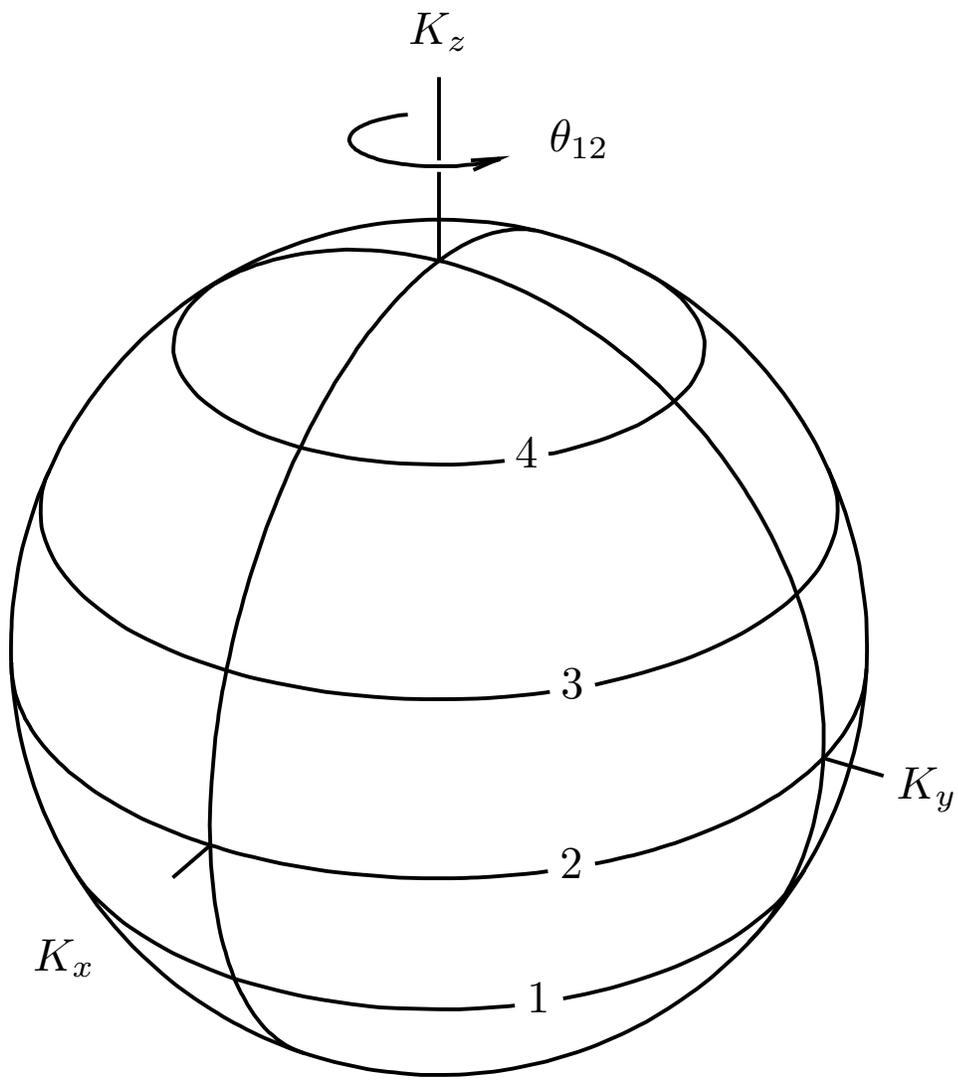
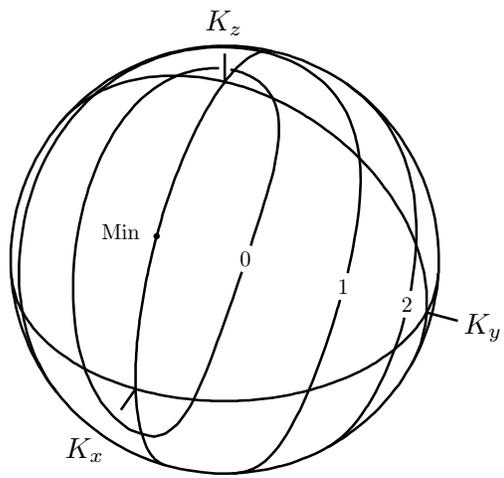
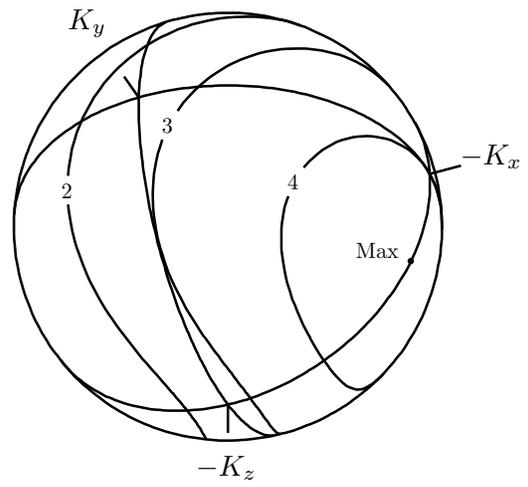
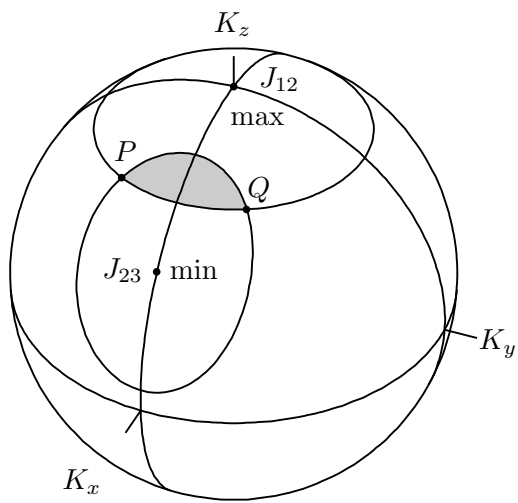
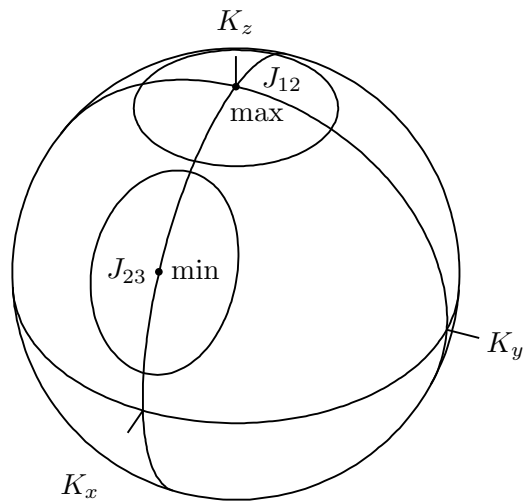
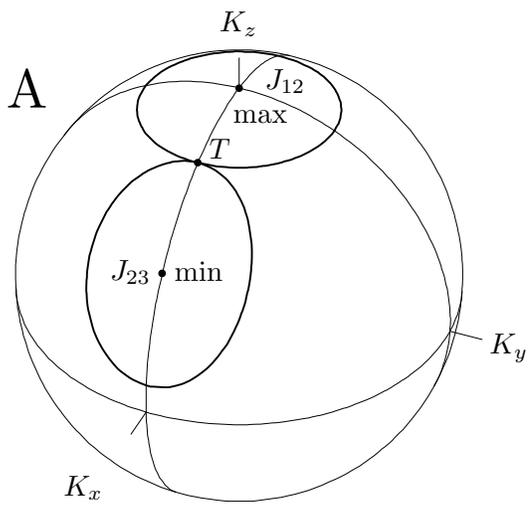
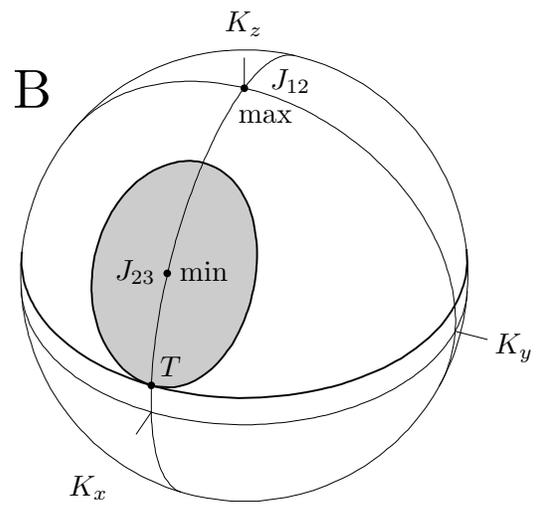
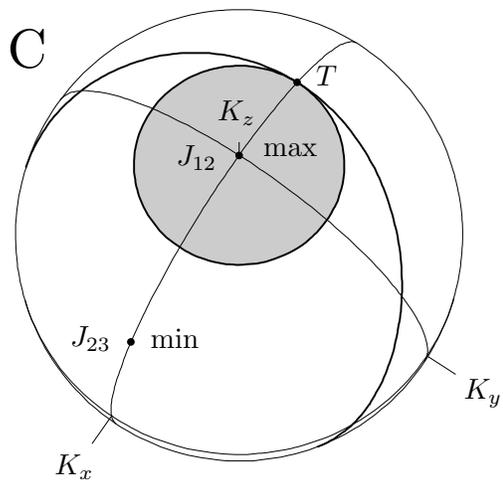
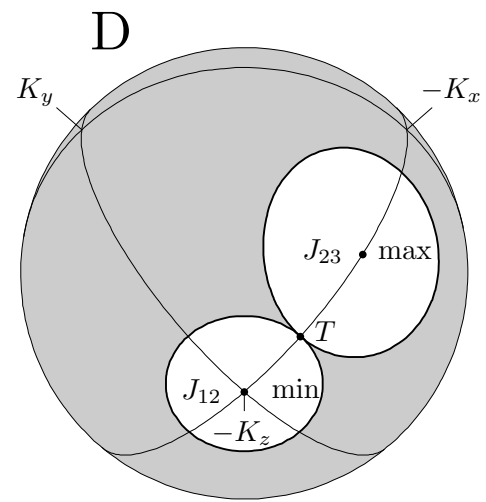
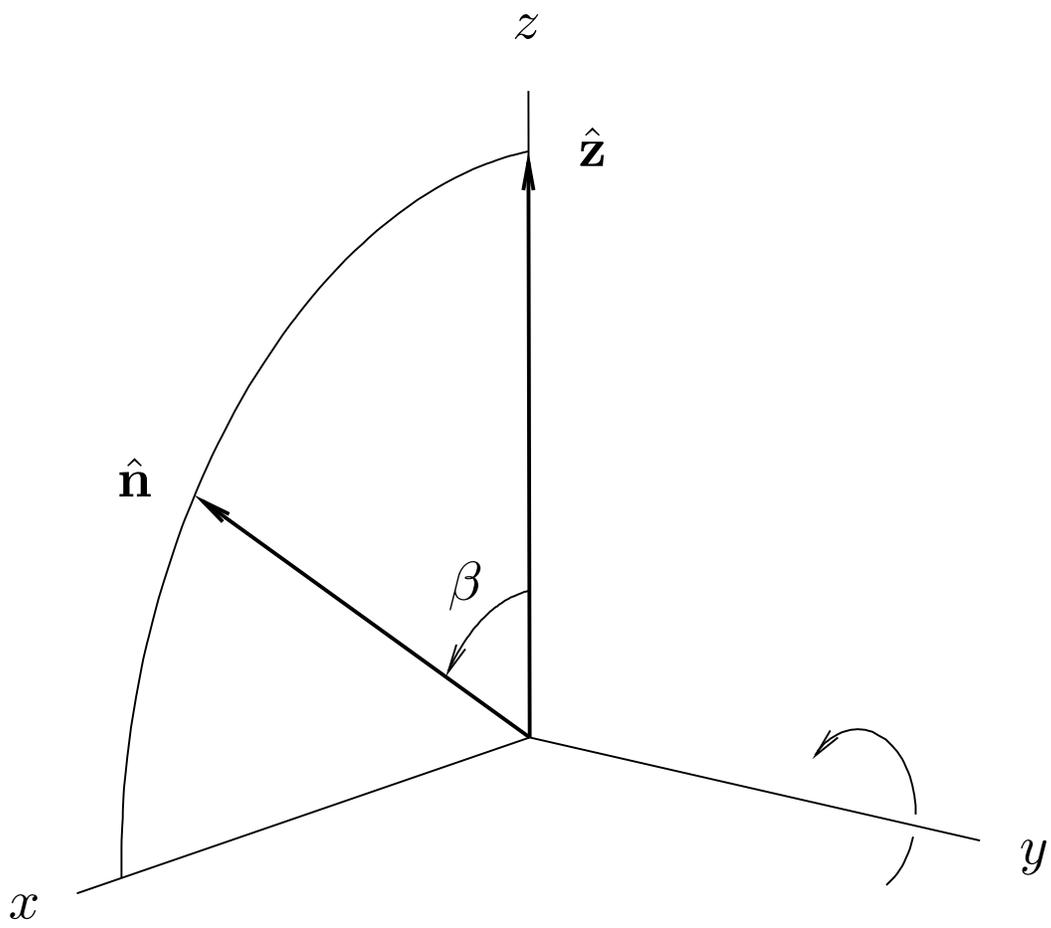
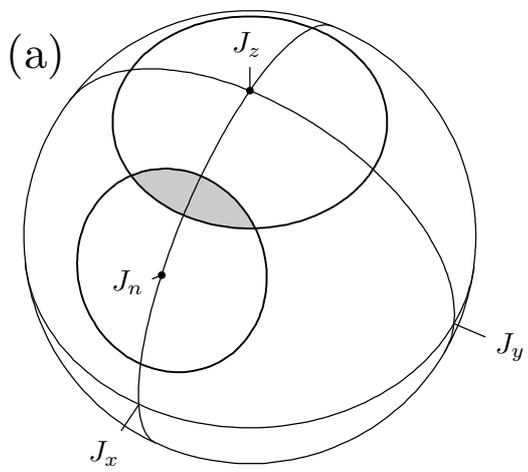
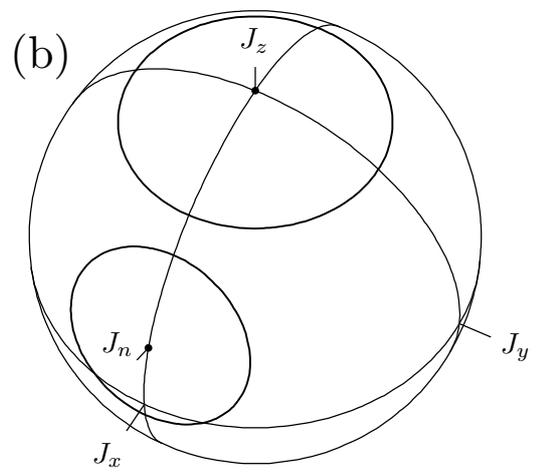
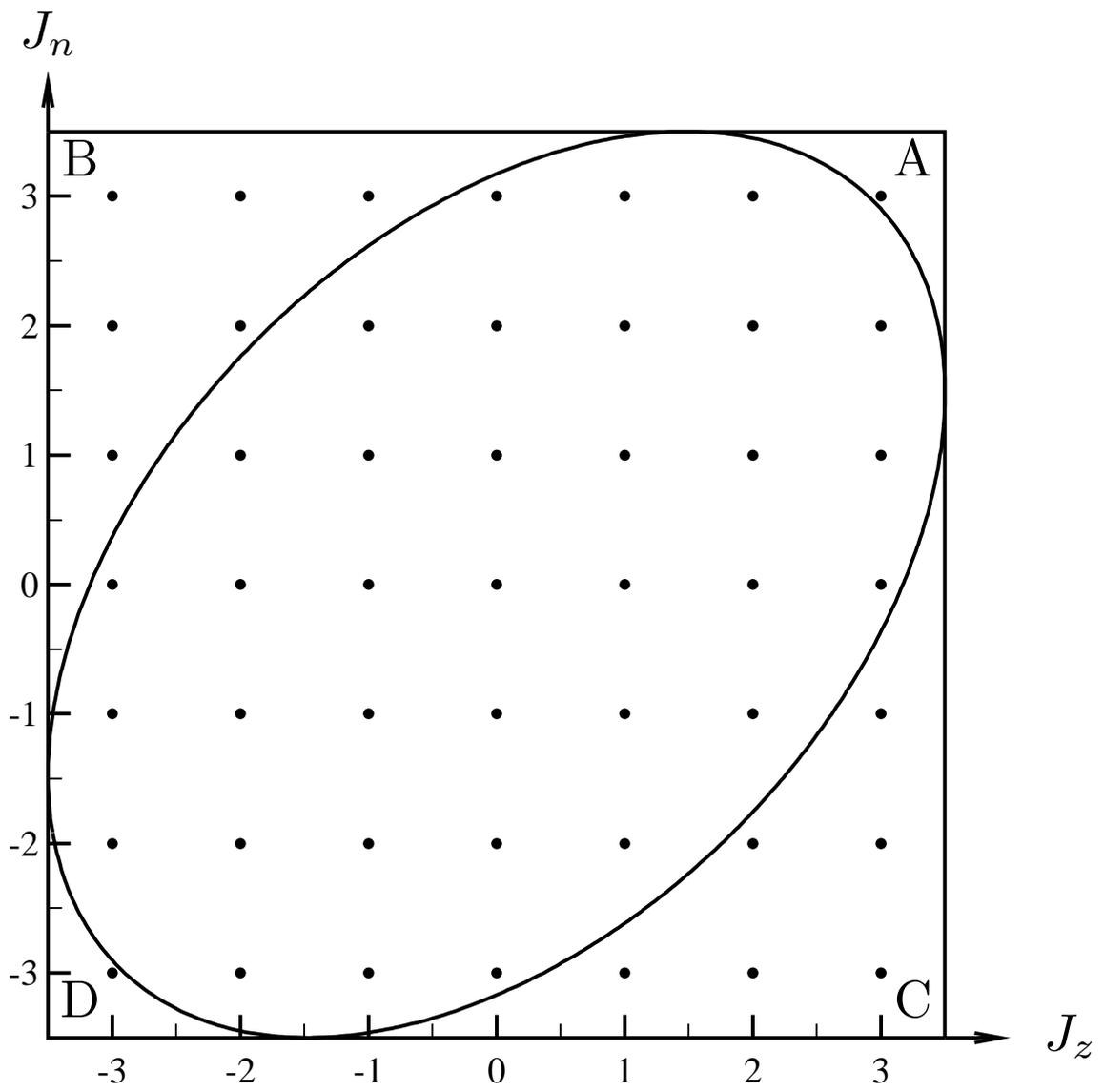
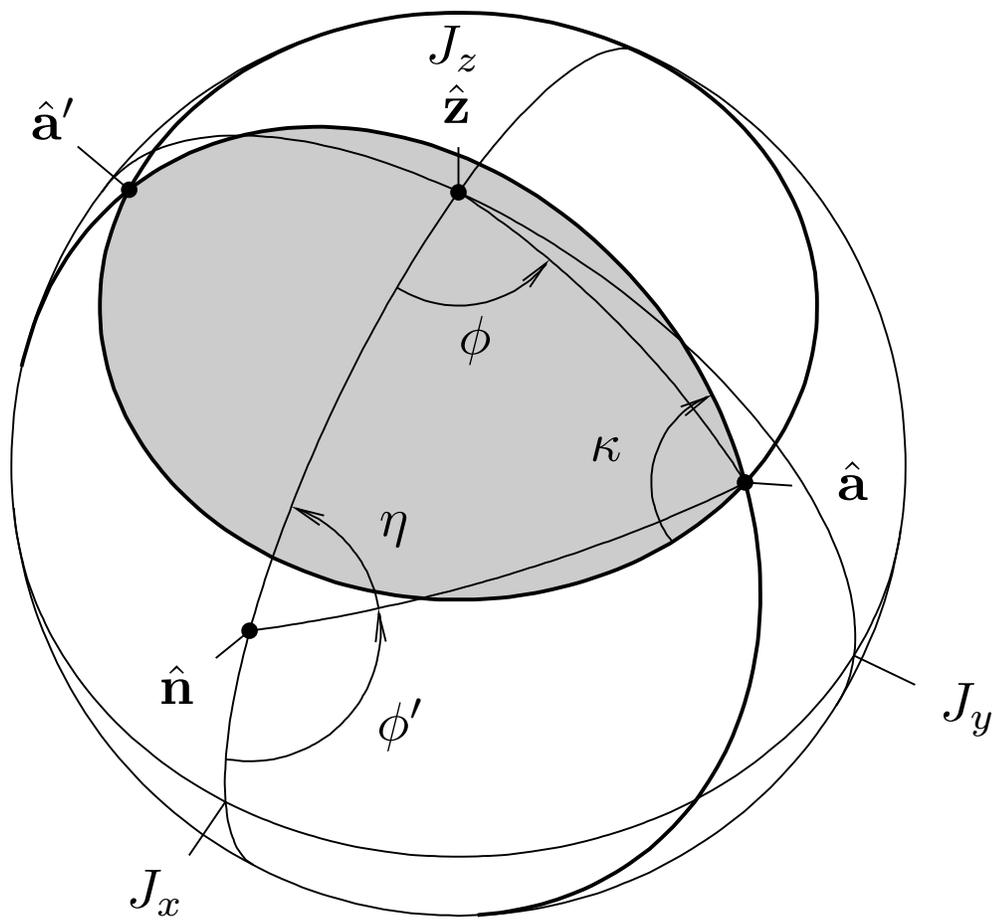
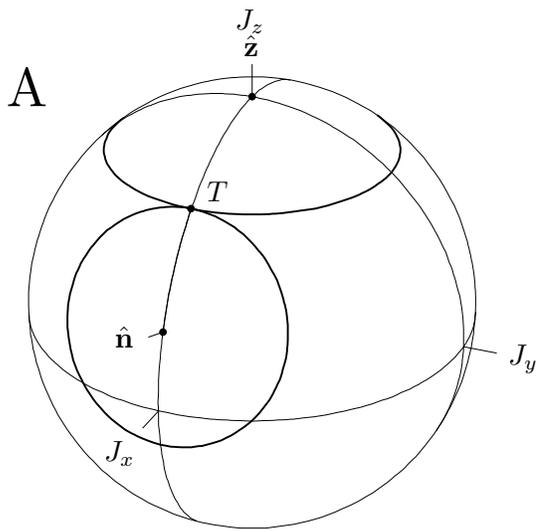
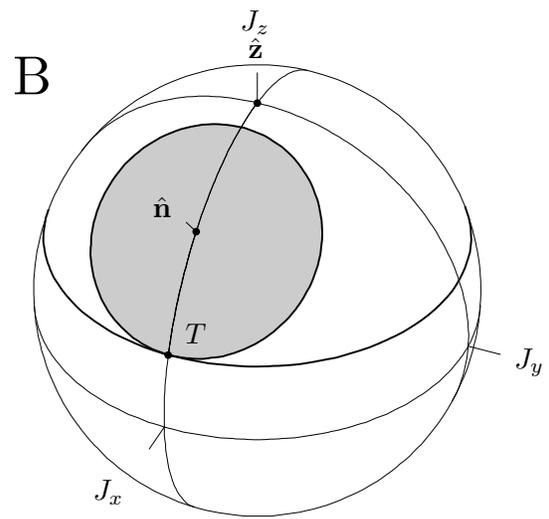
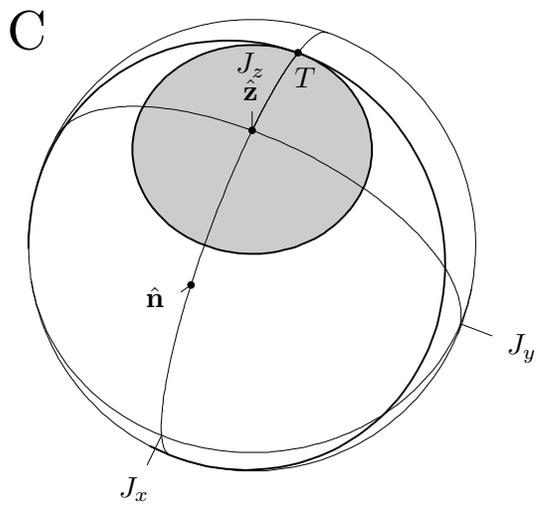
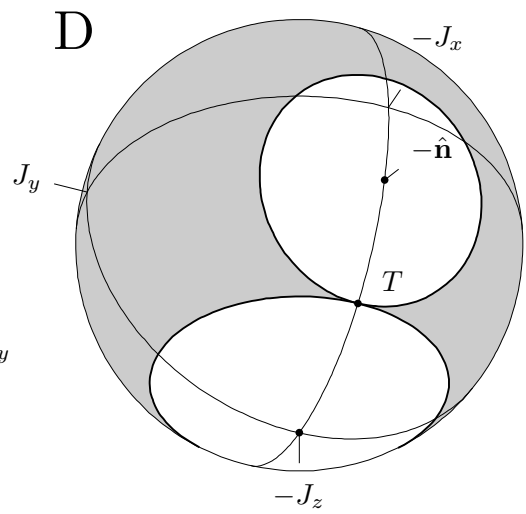
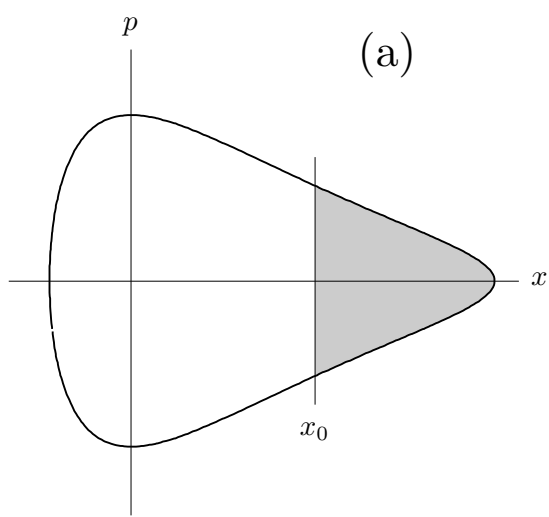
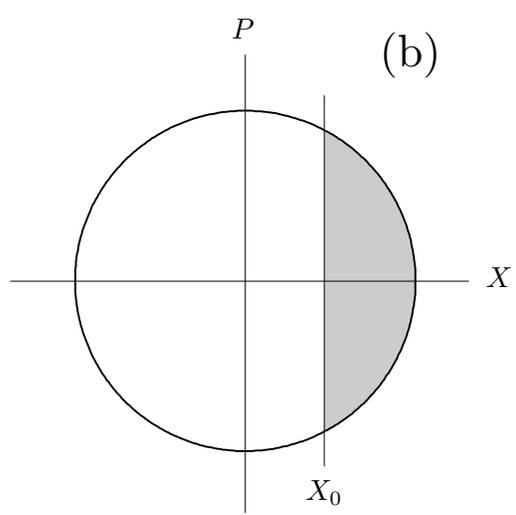
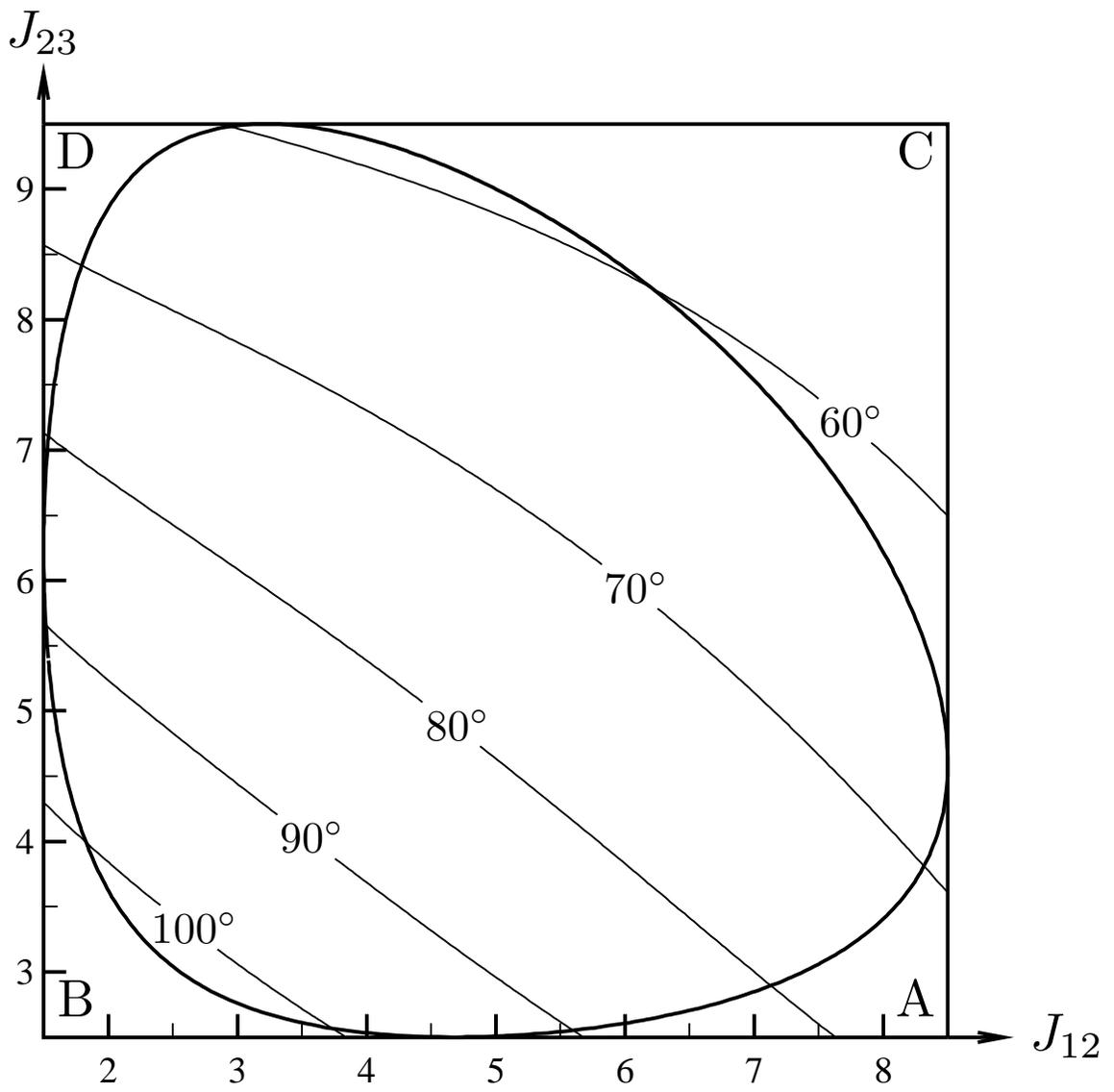
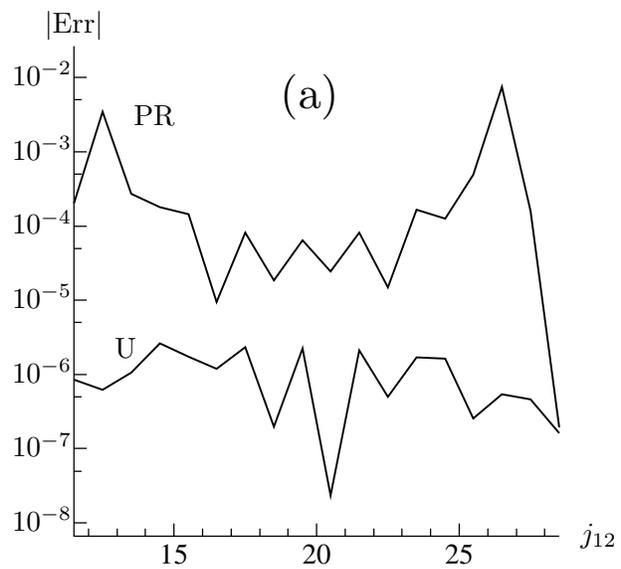
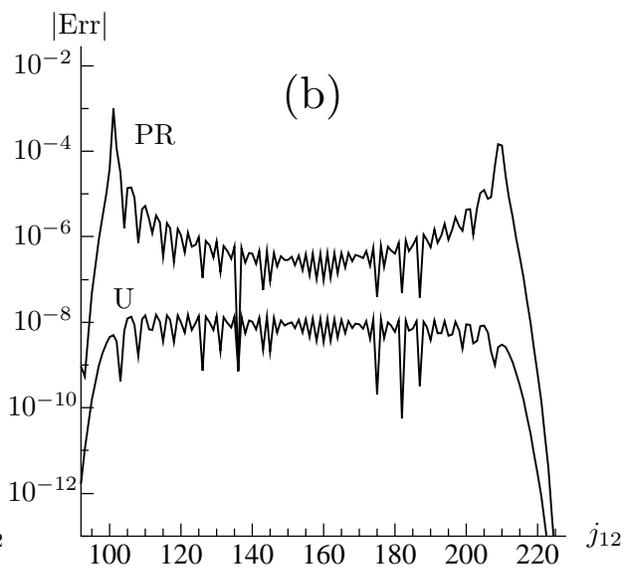
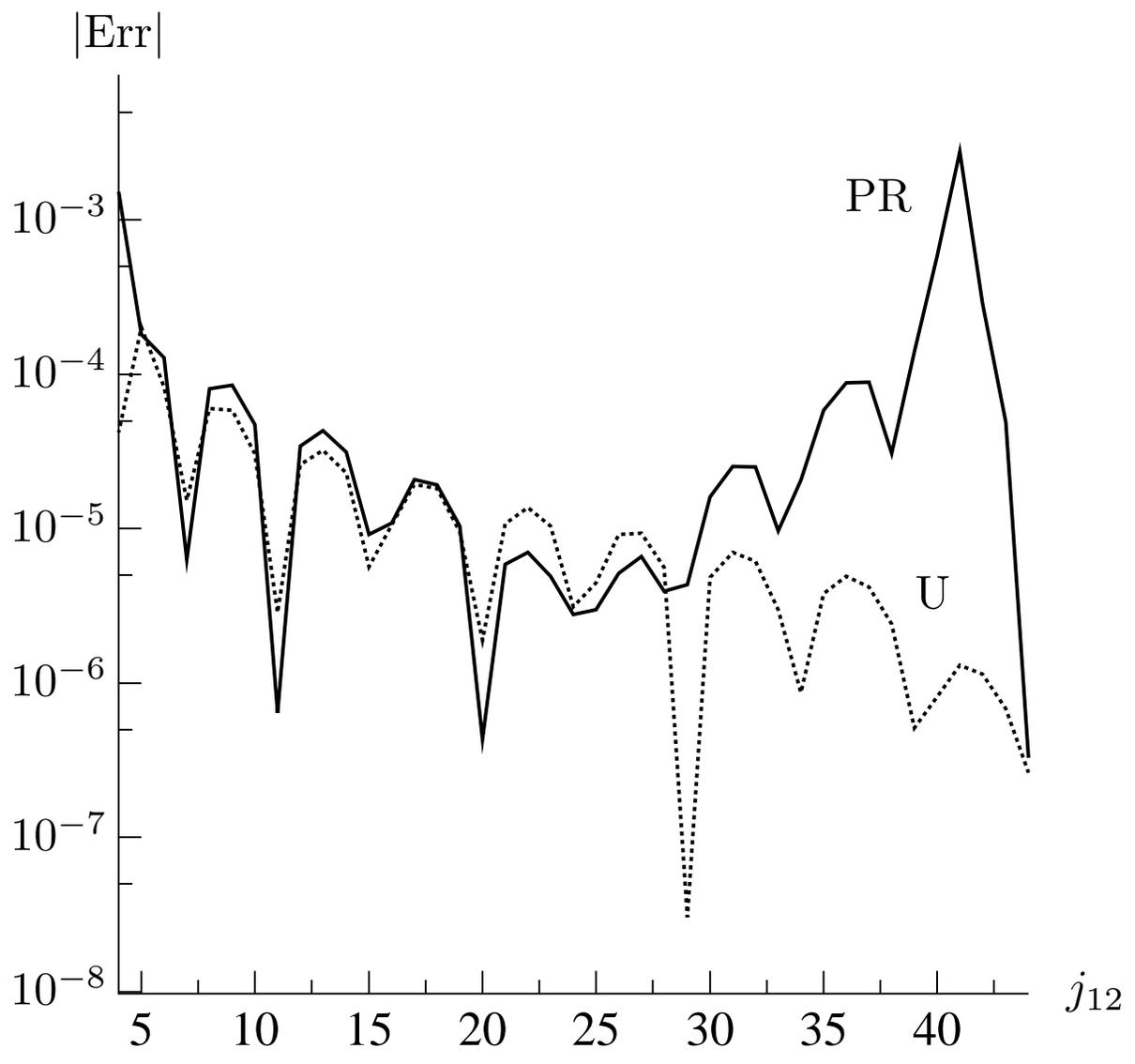
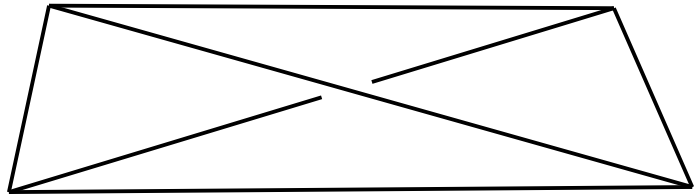
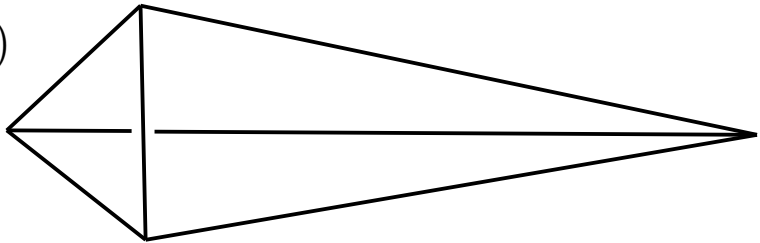
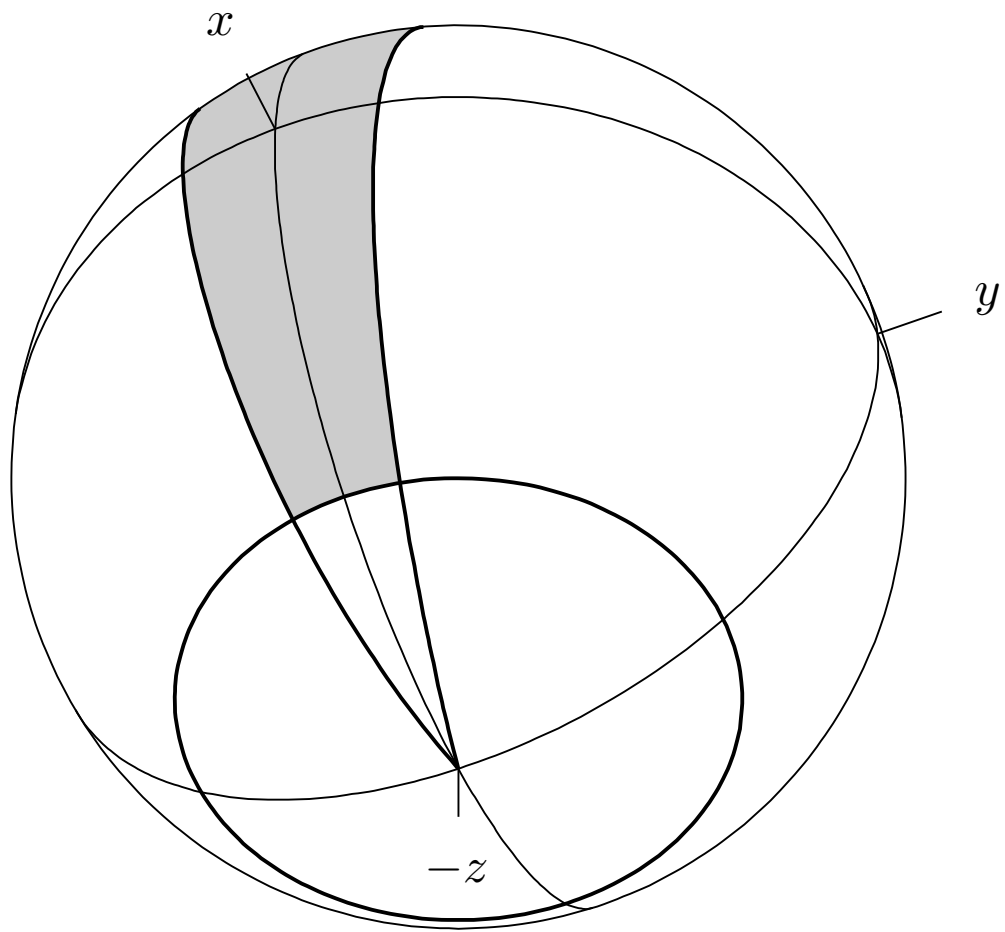
\end{document}